\documentclass[prb,aps,notitlepage]{revtex4-1}
\usepackage{graphicx} 
\newcommand{\sign}{\mathrm{sign}} 
\newcommand{\tr}{\mathrm{Tr}} 
\usepackage{color} 
\usepackage{amsbsy}

\begin{document}
\title{Kondo screening by the surface modes of a strong topological
  insulator} 
\author{E. Orignac} 
\affiliation{Laboratoire de Physique de l' \'Ecole Normale
  Sup\'erieure de Lyon, CNRS UMR5672, 46 All\'ee d'Italie, 
F-69364 Lyon Cedex 7, France} 
\author{S. Burdin} 
\affiliation{Univ Bordeaux, LOMA, UMR 5798, 
F-33400 Talence, France} 
\affiliation{CNRS, LOMA, UMR 5798, 
F-33400 Talence, France} 
\date{\today} 

\begin{abstract}
  We consider a magnetic impurity deposited on the surface of a strong
  topological insulator and interacting with the surface modes by a Kondo 
  exchange interaction. Taking into account the warping of the Fermi
  line of the surface modes, we derive a mapping to an effective one
  dimensional model and show that the impurity is fully
  screened by the surface electrons except when the Fermi level lies
  exactly at the Dirac point. Using an Abrikosov fermion mean-field
  theory, we  calculate the shape of the electronic density 
  Friedel oscillation resulting from the presence of the Kondo
  screening cloud. We analyze quantitatively the observability of a six-fold symmetry in 
the Friedel oscillations for two prototype compounds: Bi$_2$Se$_3$ and Bi$_2$Te$_3$. 
\end{abstract}

\maketitle 

\section{Introduction}
\label{sec:intro}

Topological insulators are a new class of materials, with an
insulating bulk and a conducting
surface.\cite{HasanMoore2010,HasanKane2010,QiZhang2010}. 
The existence of four topological invariants guarantees the stability
of these surface states against perturbations.\cite{Fu2007,moore2007,roy2009}  
Furthermore, the topological invariants permit to distinguish two
types of topological insulators, the weak topological insulators and
the strong topological insulators. 
In a strong topological insulator,
the surface modes at low energy form an odd number of Dirac cones.
This is to be contrasted with a strictly two-dimensional conductor
such as  graphene\cite{castroneto2009}, where the Nielsen-Ninomiya
theorem\cite{Nielsen1981} only permits an even number of Dirac cones:
the presence of the gapped bulk is crucial to the formation of an odd
number of Dirac cones. Experimentally, 
the material $\mathrm{Bi_xSb_{1-x}}$ is a strong topological insulator with 
three Dirac cones\cite{hsieh2008} while 
$\mathrm{Bi_2Se_3}$, $\mathrm{Bi_2Te_3}$ and $\mathrm{Sb_{2}Te_{3}}$, are strong topological
insulators with a single Dirac cone.\cite{zhang2009,chen2009} 
In the case of  $\mathrm{Bi_2Te_3}$, a significant hexagonal Fermi line 
warping is present.\cite{chen2009} 
More recent examples of topological insulator materials are
$\mathrm{TlBiSe_2}$\cite{sato2010}, strained HgTe\cite{bruene2011,bouvier2011}, $\mathrm{Bi_2Te_{1.6}S_{1.4}}$\cite{ji2012}
$\mathrm{PbBi_2Te_4}$\cite{kuroda2012},
$\mathrm{Pb}({\mathrm{Bi}}_{1-x}{\mathrm{Sb}}_{x}{)}_{2}{\mathrm{Te}}_{4}$\cite{souma2012},
$\mathrm{Bi_2Se_2Te}$, $\mathrm{Bi_2Te_2Se}$, and
$\mathrm{GeBi_2Te_4}$.\cite{neupane2012}
It has also recently been proposed on the basis of band structure
calculations that the ternary compound
LiAuSe\cite{zhang2011}  and the cerium filled
skutterudites\cite{yan2012}  
$\mathrm{CeOs_4As_{12}}$ and $\mathrm{CeOs_4As_{12}}$, 
 could be topological insulators. A recent review of  known
 topological insulators can be found in Ref.~\onlinecite{yan2012a}. 
In all these systems, a strong spin-orbit coupling is present, and in
the 
surface states the helicity, \textit{i.e.}, the
sign of the spin projection of the quasiparticle spin on the
quasi-momentum is fixed.   
An interesting theoretical question is then whether the Kondo effect\cite{Kondo1964,nozieres_shift,Wilson1975,andrei_kondo_review} in the
surface modes is affected by a fixed helicity. In particular, one
would like to know whether a conventional Kondo effect takes place, or
whether the fixed helicity constraint gives rise to unconventional
Kondo fixed points. 
In the case of two-dimensional
topological insulators\cite{Wu2006}, the question was
addressed\cite{tanaka2011}, and it was shown that a conventional Kondo
effect would take place, leading to a suppression of the
backscattering of the edge states by magnetic impurities. In the case of a
three-dimensional topological insulator, 
the Anderson model was considered both in the
absence\cite{Feng2009,zitko2010,tran2010} and in the presence\cite{mitchell2012} of Fermi surface warping.  
Within a variational trial wavefunction method, it was found that the local moment would be fully quenched, but
correlations would exist between the conduction electron spin and the
spin of the local impurity.\cite{Feng2009} Analogous results were
obtained in a two-dimensional electron gas with Rashba and Dresselhaus
spin-orbit couplings.\cite{feng2011} In the case of a pure Dirac
spectrum, a mapping on the one-dimensional Anderson model with a
pseudogap in the hybridization function was obtained.\cite{zitko2010,tran2010} 
It was concluded that away from the Dirac point the Kondo effect would
take place, and the impurity would be fully screened, while at the
Dirac point, the local moment would decouple.\cite{zitko2010} 
The local density of states (LDOS), the local spin density of
states (LSDOS) and the Friedel oscillations were also
investigated\cite{tran2010} within 
a U(1) slave-boson\cite{hewson1997} mean-field theory. 
In Ref.\onlinecite{mitchell2012}, the Anderson model in a topological
insulator with a Fermi surface warping was considered using the
numerical renormalization group and the behavior of the LDOS was
obtained.    
Experimentally,
magnetic impurities such as manganese\cite{noh2011}, 
nickel, iron\cite{wray2011,scholz2012,shelford2012}, cobalt\cite{ye2012,shelford2012} and
gadolinium\cite{valla2012} have been deposited on the surface 
of topological insulators. It was found that the surface states were
remarkably insensitive to the presence of both magnetic and
nonmagnetic impurities\cite{noh2011,valla2012}. While the first result can be
understood as a consequence of the suppression of backscattering, the
second result is more surprising since magnetic impurities permit
backscattering by flipping the electron spin. 

Besides this single
impurity behavior, it has been suggested theoretically that in a Kondo
lattice at electronic half-filling, the Kondo interaction could help the formation of a
topological insulator.\cite{dzero2012,dzero2012a,tran2012,zhang2012} 
There are indeed recent experimental indications that the
Kondo insulator\cite{zhang_smb6_2012,botimer2012} 
$\mathrm{SmB_6}$ could be a topological insulator. 
This also lends support to the hypothesis that Kondo screening of magnetic
impurities is compatible with the helical character of the surface
states. 
An important technique to probe conducting surfaces is  Scanning
Tunneling Microscopy (STM).\cite{binnig1982,binnig1987} This is particularly interesting in relation to the Kondo effect since
STM measurements of the LDOS around a Kondo impurity located on the surface of a metal have
already been
performed\cite{madhavan1998,madhavan2001,knorr2002,wahl2005,ysfu2007}, 
and the influence of the Kondo resonance on the LDOS has been studied theoretically\cite{ujsaghy2000}.   
Since the surface of a topological insulator is conducting, it can be
probed by  STM.\cite{roushan2009,beidenkopf2011,alpichshev2011,alpichshev2012,cheng2012,teague2012,zhang2013} 
The existing predictions for the LDOS caused by a Kondo impurity\cite{mitchell2012} could thus be tested in
that manner.  
Moreover, following the proposal of Ref.~\onlinecite{affleck2008}, 
integrating the measured LDOS would permit the measurement of the
Friedel oscillations of the electron density induced by the Kondo
screening cloud around the magnetic impurity.    

In the present paper, we want to further analyze the Kondo effect of a
magnetic impurity at the surface of a strong topological insulator
with warping. In the first part, we reduce the Kondo Hamiltonian to a
one-dimensional model which can be treated by integrability
techniques. We find that a conventional Kondo effect is obtained, with
the impurity screened by the surface modes unless the Fermi level is right
at the Dirac point, in which case, because of the vanishing density of
states, the impurity remains unscreened for weak Kondo coupling.  In
the second part, we calculate within an Abrikosov fermion mean-field
theory\cite{abrikosov_kondo} the Friedel
oscillations\cite{affleck2008} 
resulting from the existence of the Kondo screening cloud. For weak
Fermi surface warping, we derive a perturbative expression of the
density Friedel oscillations.  
In the third part we discuss the observability in STM measurements of the
Friedel oscillations of electron density\cite{affleck2008} 
around an impurity  
in the specific cases of two prototype compounds: $\mathrm{Bi_2Se_3}$
and $\mathrm{Bi_2Te_3}$.

\section{Mapping to a one-dimensional model}\label{sec:mapping}  
The free electrons Hamiltonian of the surface modes of a strong topological insulator
with warping reads:
\begin{eqnarray}\label{eq:warping} 
  H_0\equiv\sum_{\mathbf{k},\alpha,\beta} 
  c^\dagger_{\mathbf{k},\alpha} [-i v_F \mathrm{\hat{z}}\cdot
  (\boldsymbol{\sigma}_{\alpha \beta}\times \mathbf{k})  +\frac{\lambda}{2} [(k_x+ik_y)^3+(k_x-ik_y)^3]\sigma^z 
  -\mu
  \delta_{\alpha\beta} ] 
  c_{\mathbf{k},\beta} 
   &\equiv& \sum_{\mathbf{k},\alpha,\beta} c^\dagger_{\mathbf{k},\alpha}
   \mathcal{H}^0_{\alpha\beta}(\mathbf{k}) c_{\mathbf{k},\beta}~, 
\end{eqnarray}
where $c_{\mathbf{k},\alpha}$ annihilates a fermion of momentum
$\mathbf{k}$ and spin $\alpha$, $\hat{z}$ is the normal to surface of
the topological insulator, $\boldsymbol{\sigma}=(\sigma^{x}, \sigma^y,\sigma^z)$ denotes the Pauli matrices,
  $v_F$ is the Fermi velocity, $\lambda$ is the warping, and $\mu$ the
  chemical potential. In the case of $\mathrm{Bi_2Se_3}$, one
  has\cite{kuroda2010} $v_F=3.55$ eV\AA~ 
and $\lambda=128$
  eV\AA$^3$. For $\mathrm{Bi_2Te_3}$, \cite{chen2009,an2012} one has
  $v_F=2.55$ eV\AA~and $\lambda=250$ eV\AA.   
In the following we will study the effects induced by the presence of a magnetic quantum impurity. 
We take the position of the impurity as the origin of our
  coordinates so that the Kondo Hamiltonian describing the surface modes and their interaction with the impurity reads: 
\begin{eqnarray}\label{eq:kondo} 
H&=&H_0+H_{K}
\equiv H_0+\frac{J_K}{L^2}
  \sum_{\mathbf{k},\mathbf{k'},\alpha,\beta}\mathbf{S}\cdot
  c^\dagger_{\mathbf{k},\alpha} \boldsymbol{\sigma}_{\alpha \beta} c_{\mathbf{k'},\beta}~,  
\end{eqnarray}   
where $J_K$ is the Kondo interaction, $L$ the linear dimension of the
(square) surface, $\mathbf{S}$ the impurity spin. 

\subsection{Eigenstates of the free Hamiltonian}
\label{sec:eigenstates}
In the present section, we briefly review the eigenstates of the free
electrons  Hamiltonian~(\ref{eq:warping}). 
We will use a spinor notation to represent the Fermion annihilation
and creation operators:
\begin{equation}
\label{eq:spinor}
\Psi(\mathbf{r})=\left(
\begin{array}{c}
\psi_\uparrow(\mathbf{r}) \\ \psi_\downarrow (\mathbf{r}) 
\end{array}
\right) = \sum_{s=\pm} \int \frac{d^2
                     \mathbf{k}}{(2\pi)^2}
                   \psi_{\mathbf{k}}^s(\mathbf{r}) c_{s}(\mathbf{k}) ,
\end{equation}
where the spinors $\psi_{\mathbf{k}}^s(\mathbf{r})$ are 
eigenstates of the first quantized Hamiltonian and 
$\{c_s(\mathbf{k}),c_{s'}(\mathbf{k'})\}= (2\pi)^2 \delta_{s,s'}
\delta(\mathbf{k}-\mathbf{k'})$.  
Introducing the polar coordinates  $\mathbf{k}\equiv (k, \phi)$ and $\mathbf{r}\equiv (r, \theta)$ 
the spinors have the explicit form: 
\begin{equation}\label{eq:eigenstates} 
\psi_{k,\phi}^{s}(r,\theta)=\left(
\begin{array}{c}
A_s \\ -i B_s e^{i \phi}
\end{array}\right) e^{i k r \cos (\theta -\phi)},  
\end{equation} 
where $|A_s|^2+|B_s|^2=1$ and 
\begin{equation}
  \label{eq:alpha-beta}
  \frac{A_s+iB_s}{A_s-iB_s}=s \frac{\lambda k^3 \cos 3 \theta +iv_F k}{\sqrt{v_F^2 k^2 + \lambda^2 k^6 \cos^2 3\theta}}~.   
\end{equation}
We note that $A_s$ and $B_s$ are periodic functions of  $\theta$
of period $\frac{2\pi} 3$.  In the rotationally invariant case, i.e., without the warping term ($\lambda=0)$, 
the eigenstates of the Hamiltonian are rewritten as angular 
momentum eigenstates, and only the s-wave
channel\cite{zitko2010,tran2010} is found to contribute to the Kondo
interaction. In the case with warping ($\lambda\neq 0$),  there is
only  a discrete rotational symmetry, and instead  of representations
of $O(2)$ the eigenstates have to be expressed as representations of the
discrete group $Z_3$:    
\begin{eqnarray}\label{eq:angular} 
  \psi^s_{k,\phi,\ell}(\mathbf{r})&=&\frac 1 {\sqrt{3}} \sum_{n=0,1,2} e^{i \frac
    {2\pi}3 \ell n} \psi^s_{k,\phi+\frac{2n\pi} 3}(\mathbf{r}),  
\end{eqnarray}
with the restriction $|\phi|<\pi/3$ and $\ell=0,1,2$ labels the
representation.  With full rotational symmetry,
$\ell$ would be the angular momentum. Because of the 3-fold symmetry of the
warping term, the states having a difference of angular momentum equal
to a multiple of 3 are hybridized together, and the  angular momentum
$\ell$  is defined only modulo 3. In that basis, the spinor~(\ref{eq:spinor}) reads: 
\begin{eqnarray}
  \Psi(\mathbf{r})= \sum_{\ell,s=\pm} 
  \int\frac{d^2\mathbf{k}}{(2\pi)^2} 
  \psi^{s}_{\mathbf{k},\ell}(\mathbf{r}) c_{\ell,s}(\mathbf{k})~. 
\end{eqnarray}
Since $\psi^{s}_{k,\phi,1}(0,\theta)=0$, only the $\ell=0,2$ modes interact with the magnetic impurity: 
   \begin{equation}\label{eq:spinor-z3} 
      \Psi(\mathbf{0})=\sqrt{3} \int\frac{d^2\mathbf{k}}{(2\pi)^2}  \left(
        \begin{array}{c}
          A c_{0,+}(\mathbf{k})  -B c_{0,-}(\mathbf{k})  \\ 
          -i B c_{2,+}(\mathbf{k})  -iA c_{2,-}(\mathbf{k})  
        \end{array}
\right)~, 
    \end{equation}
and the Kondo Hamiltonian is rewritten as: 
 \begin{eqnarray}\label{eq:hamiltonian-polar} 
      H&=&\sum_{\ell,s} \int\frac{d^2\mathbf{k}}{(2\pi)^2} (s
      E(\mathbf{k})-\mu)  c^\dagger_{\ell,s}(\mathbf{k})
      c_{\ell,s}(\mathbf{k}) + J_K \mathbf{S} \cdot \Psi^\dagger(\mathbf{0})
       \boldsymbol{\sigma}\Psi(\mathbf{0})~, 
\end{eqnarray}
with 
\begin{eqnarray}
  E(\mathbf{k})\equiv \sqrt{v_F^2 k^2 +\lambda^2 k^6 \cos^2(3\phi)}~. 
    \end{eqnarray}
The Hamiltonian~(\ref{eq:hamiltonian-polar}) can be further reduced by
turning the integration variable $\mathbf{k}$ to a system of curvilinear coordinates 
 $(E,\kappa_\perp)$, where $E$ is the energy of the eigenstate, and
 $\kappa_\perp$ is the curvilinear coordinate along the line of
 constant energy. Introducing the new operators: 
\begin{eqnarray}
a_{\ell,s}(E,\kappa_\perp)=\frac{c_{\ell,s}(\mathbf{k})}{\sqrt{||\nabla_{\mathbf{k}}
          E(\mathbf{k})||} }~,
    \end{eqnarray}
with anticommutation relations $ \{a_{\ell,s}(E,\kappa_\perp),a^\dagger_{\ell,s}(E',\kappa'_\perp)\}=(2\pi)^2
      \delta(E-E') \delta(\kappa_\perp-\kappa'_\perp)$, the free electrons part of the Hamiltonian~(\ref{eq:hamiltonian-polar}) can be rewritten:      
\begin{eqnarray} 
H_0=\sum_{\ell,s} \int \frac{dE d\kappa_\perp}{(2\pi)^2} (sE-\mu)
    a^\dagger_{\ell,s}(E,\kappa_\perp) a_{\ell,s}(E,\kappa_\perp)~. 
\end{eqnarray}
Introducing the density of states:
\begin{eqnarray} 
\rho_0(E) \equiv\int \frac{d
        \kappa_\perp}{4\pi^2 ||\nabla_{\mathbf{k}}
          E(\mathbf{k})|| }~, 
\end{eqnarray} 
and the operators: 
\begin{eqnarray} 
a_{\uparrow,+}(E)=\sqrt{3}\int d\kappa_\perp\frac{ A(E,\kappa_\perp)
  a_{0,+}(E,\kappa_\perp) }{ \sqrt{2\pi^2 \rho_0(E)
    ||\nabla_{\mathbf{k}}
          E(\mathbf{k})||}}~,  \\ 
a_{\uparrow,-}(E)= \sqrt{3}\int d\kappa_\perp \frac{B(E,\kappa_\perp) a_{0,-}(E,\kappa_\perp)}{ \sqrt{2\pi^2 \rho_0(E)
    ||\nabla_{\mathbf{k}}
          E(\mathbf{k})||}}~,   \\
a_{\downarrow,+}(E)=\sqrt{3}\int d\kappa_\perp  \frac{B(E,\kappa_\perp)
  a_{2,+}(E,\kappa_\perp) }{ \sqrt{2\pi^2 \rho_0(E)
    ||\nabla_{\mathbf{k}}
          E(\mathbf{k})||}}~, \\
a_{\downarrow,-}(E)= \sqrt{3}\int d\kappa_\perp \frac{A(E,\kappa_\perp)
  a_{2,-}(E,\kappa_\perp)}{ \sqrt{2\pi^2 \rho_0(E)
    ||\nabla_{\mathbf{k}}
          E(\mathbf{k})||}}~, 
    \end{eqnarray}
with anticommutators 
$\{a_{\alpha,s}(E),a^\dagger_{\alpha',s'}(E')\}=(2\pi)^2 \delta(E-E')
\delta_{ss'} \delta_{\alpha \alpha'}$ we can define:  
   \begin{eqnarray}\label{eq:kondo-s-wave} 
     &&  a_\alpha(E)=(-i)^{1/2-\alpha}\sum_{s=\pm}\Theta(sE)
      s^{1/2+\alpha} a_{\alpha,s}(|E|)~,  \\ 
     && \psi_\alpha(0)= \int_{-\infty}^{\infty} \frac{dE
        \sqrt{\rho_0(|E|)}}{\pi \sqrt{8}} a_\alpha (E)~,   
\end{eqnarray}
where $\Theta$ denotes the Heaviside function. 
In terms of these operators, the
Kondo Hamiltonian~(\ref{eq:hamiltonian-polar}) reads:  
\begin{eqnarray}\label{eq:reduced-kondo}  
H=J_K \mathbf{S} \cdot \Psi^\dagger(\mathbf{0})
       \boldsymbol{\sigma}\Psi(\mathbf{0})
+\int_{-\infty}^{+\infty} \frac{dE}{(2\pi)^2}  \sum_\alpha (E-\mu)
      a^\dagger_\alpha (E) a_\alpha(E) +\ldots~,       
    \end{eqnarray}
where $\ldots$ stands for the modes of the free electrons Hamiltonian $H_0$ that do
not couple to the impurity. 
Away from the Dirac point,  the density of states $\rho_0(E)$ can be
approximated by the density of states at the Fermi level $\rho_0(\mu)$, and the
usual single channel Kondo problem is obtained. For a spin-1/2
impurity,  The Kondo temperature can be obtained from the Bethe Ansatz
solution as\cite{andrei_trieste93}:
\begin{equation}
  \label{eq:kondo-temp}
   T_K = D
    e^{\gamma_E-1/4} \exp\left(\frac{-1}{J_K \rho_0(\mu)}\right),  
\end{equation}
where $\gamma_E\simeq 0.577$ is the Euler-Mascheroni constant and $D$
is a symmetric cutoff around the Fermi energy. The density of states
$\rho_0(E)$ can be expressed in terms of elliptic
integrals\cite{adroguer2012}, so the full dependence of the Kondo
temperature on the chemical potential is known up to the prefactor
$D$.  
Since the Bethe Ansatz
solution of the Kondo problem\cite{andrei_trieste93} requires a
constant density of states, $D$ represents as the energy scale away
from the Fermi energy for which the density of states starts to
deviate significantly from the density of states at the Fermi energy.    
For small warping $\lambda$ or not too far from the Dirac point, the density
of states is a linear function of energy, and $D\simeq \mu$, so the
dependence of $D$ on $E$ is only a subdominant contribution.   
Close to the Dirac point, $\mu \to 0$, the density of states $\rho_0(\mu)\sim
 |\mu|$. Because of such pseudogap, the Kondo
temperature vanishes\cite{Withoff1990}. These results are in
agreement with the ones derived in the framework of the Anderson
model\cite{zitko2010,tran2010} where Kondo screening was obtained only 
when the density of states at the Fermi level was non-vanishing.  
In the case of an impurity with spin $S>1/2$
Eq.~(\ref{eq:reduced-kondo}) would give the underscreened single
channel Kondo fixed point. 
 
More generally,  it can be established in any dimension $d$ that with any free Hamiltonian of the form: 
\begin{equation}
  \label{eq:dirac-like}
  H_0=\sum_{\alpha,\beta} c^\dagger_{k,\alpha} 
(\mu\delta_{\alpha\beta}+\boldsymbol{\epsilon}(\mathbf{k})\cdot \boldsymbol{\sigma}_{\alpha\beta}) 
c_{k,\beta}~, 
\end{equation}
having time reversal symmetry (i. e. $\boldsymbol{\epsilon}(-\mathbf{k})=-\boldsymbol{\epsilon}(\mathbf{k})$) only a conventional Kondo effect can be obtained. 
Indeed, if we write the partition function as\cite{anderson_kondo_2}: 
\begin{equation}
  \label{eq:kondo-partition}
  Z=Z_0\left\langle T_\tau e^{-J_K \int_0^{\beta}d\tau \mathbf{S}(\tau)\cdot
      \psi^\dagger_\alpha(0,\tau)\boldsymbol{\sigma}_{\alpha \beta} \psi_\beta(0,\tau) }\right \rangle_{H_0}~, 
\end{equation}
with $Z_0\equiv Tr~e^{-\beta H_0}$, 
and expand in powers of $J_K$, the series will depend on:
\begin{equation}
  \label{eq:green-general}
  G_{\alpha\beta}(0,\tau)=\frac{1}{\beta} \sum_{i\nu_n} \int \frac{d^d \mathbf{k}}{(2\pi)^d} \frac{(i\nu_n+\mu)\delta_{\alpha \beta} +\boldsymbol{\epsilon}(\mathbf{k})\cdot \boldsymbol{\sigma}_{\alpha\beta}}{(i\nu_n+\mu)^2 -
\Vert\boldsymbol{\epsilon}(\mathbf{k})\Vert^2} e^{i\nu_n \tau}~, 
\end{equation}
but since $\boldsymbol{\epsilon}(\mathbf{k})$ is odd, introducing the density of states 
$\rho_0(E)=\int \delta(E-\Vert\boldsymbol{\epsilon}(\mathbf{k})\Vert ) d^d\mathbf{k}/(2\pi)^d$ we can write:
\begin{equation}
   G_{\alpha\beta}(0,\tau)=\frac{1}{2\beta} \sum_{i\nu_n} \int \frac{\rho_0(|E|)\delta_{\alpha\beta}}{i\nu_n +\mu -E} dE~,
\label{eq:green-general2}   
\end{equation}
showing that the partition function is the same as the one 
of a system without spin-orbit coupling  having the same density of
states $\rho_0$ as the
Hamiltonian~(\ref{eq:dirac-like}). 
As a result, a conventional Kondo effect is realized every time the density of states at the Fermi level is nonzero\cite{anderson_kondo_2}. As a consequence, the dispersion and the Kondo self-energy are expected to 
remain spin-independent much beyond the mean-field approximation that we will consider in the following.

\section{Friedel oscillations and Abrikosov fermions mean-field theory}\label{sec:friedel} 

\subsection{Mean-field theory}
\label{sec:abrikosov}

\subsubsection{Abrikosov fermions and mean-field self-consistent relations}
We have seen in Sec.\ref{sec:mapping} that even in the presence of
warping, a magnetic impurity on the surface of a topological insulator
is always screened provided the density of states at the Fermi level
is nonzero. In such conditions, a 
Kondo screening cloud\cite{affleck2008} is formed around the impurity,
and Friedel oscillations are formed. We consider in the
following the case of a fully screened $S=1/2$ impurity. Since we are
dealing with a conventional Kondo fixed point, we can use the  
Abrikosov fermion representation\cite{abrikosov_kondo} for the localized spin: 
\begin{eqnarray}\label{eq:abrikosov} 
    S^+=f^\dagger_\uparrow f_\downarrow ;~~
    S^-=f^\dagger_\downarrow f_\uparrow ;~~
    S^z=\frac 1 2 ( f^\dagger_\uparrow f_\uparrow -
    f^\dagger_\downarrow f_\downarrow)~,  
\end{eqnarray}
with the constraint 
\begin{eqnarray}
\label{eq:constraint} 
    1&=& f^\dagger_\uparrow f_\uparrow + 
    f^\dagger_\downarrow f_\downarrow~, 
  \end{eqnarray}
to rewrite the Kondo interaction as a local fermion-fermion
interaction:  
\begin{eqnarray}\label{eq:4fermions} 
H_K= 
J_K \mathbf{S} \cdot \Psi^\dagger(\mathbf{0})
       \boldsymbol{\sigma}\Psi(\mathbf{0})
= \frac{J_K}2 \left( \sum_{\alpha\beta}
  f^\dagger_\alpha f_\beta\psi^\dagger_\beta(0) \psi_\alpha(0) 
-\sum_{\alpha}\frac{\psi^\dagger_\alpha(0) \psi_\alpha(0)}{2}\right)~,   
\end{eqnarray}
where the two terms on the r.h.s. respectively correspond to the spin-flip and charge potential scattering processes 
of conduction electrons on the Kondo impurity. Hereafter we will concentrate on the spin-flip interaction term and we will neglect the charge potential scattering one. 
Adding a Lagrange multiplier $\mu_f(1- \sum_\alpha f^\dagger_\alpha f_\alpha )$ 
 to the full Kondo Hamiltonian $H_0+H_K$ to enforce the constraint
 (\ref{eq:constraint}),  we  
 decouple~(\ref{eq:4fermions}) by a mean-field approximation: 
\begin{eqnarray}\label{eq:meanfield-ham} 
H_{MF}&=&H_0+ \Delta
  \sum_\alpha (f^\dagger_\alpha \psi_{\alpha}(0) +
  \mathrm{H. c.}) -\mu_f \sum_\alpha f^\dagger_\alpha f_\alpha, 
\end{eqnarray}
where the effective hybridization $\Delta$ and the Lagrange multiplier $\mu_f$ satisfy the self-consistent relations :
\begin{eqnarray}
\label{eq:meanfield-scDelta} 
\Delta &=& \frac{J_K}{2} \sum_\alpha \langle \psi^\dagger_\alpha f_\alpha \rangle~, \\
\label{eq:meanfield-scmuf} 
1&=& \sum_\alpha \langle f^\dagger_\alpha f_\alpha \rangle~,   
\end{eqnarray} 
where $\langle\cdots\rangle$ denotes the thermal average computed with the mean-field 
effective Hamiltonian~(\ref{eq:meanfield-ham}). 
We introduce the Fourier decomposition:
\begin{equation}
  \label{eq:fourier-operators}
  \psi_\alpha(\mathbf{r}) = \frac 1 {L} \sum_{\mathbf{k}}
  c_{\mathbf{k}\alpha} e^{i \mathbf{k} \cdot \mathbf{r}}~, 
\end{equation}
where $L^2$ is the surface of the system. 
To solve the mean field equations, we introduce the Green's functions 
\begin{eqnarray}
  \label{eq:green-cf}
  G_{\alpha\beta}^{cc}(\mathbf{k},\mathbf{k'},\tau)&=&-\langle T_\tau c_{\mathbf{k}\alpha}(\tau)
  c_{\mathbf{k'}\beta}^{\dagger}(0)\rangle~, \nonumber \\ 
  G_{\alpha\beta}^{fc}(\mathbf{k'},\tau)&=&-\langle T_\tau f_{\alpha}(\tau)
  c_{\mathbf{k'}\beta}^{\dagger}(0)\rangle~, \nonumber \\
   G_{\alpha\beta}^{cf}(\mathbf{k},\tau)&=&-\langle T_\tau
   c_{\mathbf{k}\alpha}(\tau) f^\dagger_{\alpha}(0) \rangle~,
   \nonumber \\ 
  G_{\alpha\beta}^{ff}(\tau)&=&-\langle T_\tau f_{\alpha}(\tau)
  f_{\beta}^{\dagger}(0)\rangle~. 
\end{eqnarray}
Using the equations of motion from the 
Hamiltonian~(\ref{eq:meanfield-ham}) and a Fourier decomposition in Matsubara frequencies, the Green's functions in (\ref{eq:green-cf}) are expressed as:  
\begin{eqnarray}
G^{cc}(\mathbf{k},\mathbf{k'},i\nu_n)&=&\delta_{\mathbf{k},\mathbf{k'}}G_{0}(\mathbf{k},i\nu_n) 
+\frac{|\Delta|^2}{L^2} G_{0}(\mathbf{k},i\nu_n)G^{ff}(i\nu_n) G_{0}(\mathbf{k'},i\nu_n)~, 
\label{eq:gcc-sol} \\
 G^{fc}(\mathbf{k},i\nu_n) &=&-\frac{\Delta^\star}L G^{ff}(i\nu_n) G_{0}(\mathbf{k},i\nu_n)~, 
\label{eq:gfc-sol} \\
G^{cf}(\mathbf{k},i\nu_n)&=&-\frac{\Delta} L G_{0}(\mathbf{k},i\nu_n)G^{ff}(i\nu_n)~, 
\label{eq:gcf-sol} \\
G^{ff}(i\nu_n)&=&(i\nu_n+\mu_f
  -\Sigma(i\nu_n))^{-1}, \label{eq:gff-sol}  
\end{eqnarray}
with the free electrons Green function :
\begin{eqnarray}
G_{0}(\mathbf{k},i\nu_n)&\equiv&
(i\nu_n -{\cal H}^0(\mathbf{k}))^{-1}~,   \label{eq:gcc-free}  
\end{eqnarray}
and the self-energy: 
\begin{eqnarray}
\Sigma(i\nu_n)=|\Delta|^2\int \frac{d^2 \mathbf{k}}{(2\pi)^2}G_{0}(\mathbf{k},i\nu_n)~. 
\label{eq:free-energy}
\end{eqnarray}
Hereafter, we introduce the non-interacting electronic density of states, 
$\rho_0(E)\equiv\int\delta(E-\epsilon_\mathbf{k})d^2\mathbf{k}/(2\pi)^2$, where 
$\epsilon_\mathbf{k}$ denotes the electronic eigenenergies. 
Remarking that ${\cal H}^0(\mathbf{k})$ has time reversal symmetry, and invoking a similar analysis as the one 
leading to Eq.~(\ref{eq:green-general2}), we find the following spin-independent expression for the 
self-energy: 
\begin{eqnarray}
  \Sigma_{\alpha\beta}(z)&\equiv&
\delta_{\alpha\beta} |\Delta|^2 \int_{-\infty}^\infty dE
\frac{\rho_0(E)}{z+\mu-E}~.   \label{eq:sigma-sol-1}  
\end{eqnarray}
Introducing the real and imaginary parts of the self-energy, 
$\Sigma_{\alpha\beta}(E+i0^+)=\delta_{\alpha\beta}[\Sigma'(E)+i\Sigma''(E)]$, with 
\begin{eqnarray}
\Sigma'(E)&=&|\Delta|^2 \int_{-\infty}^\infty d\epsilon\rho_0(\epsilon)
\mathrm{P.V.}\left( \frac{1}{E+\mu-\epsilon}\right)~, 
\label{eq:Sigmaprime}\\
\Sigma''(E)&=&-\pi|\Delta|^2\rho_0(E+\mu)~, 
\label{eq:Sigmasecond}
\end{eqnarray}
the self-consistency conditions~(\ref{eq:meanfield-scDelta}) and~(\ref{eq:meanfield-scmuf}) read: 
\begin{eqnarray}\label{eq:mf-general} 
|\Delta|^2 &=&  J_K  \int_{-\infty}^\infty \frac{dE}{\pi} n_F(E)
  \frac{(E +\mu_f) \Sigma''(E)}{[E + \mu_f
    -\Sigma'(E)]^2 + [\Sigma''(E)]^2}~, \\
\label{eq:mf-generalbis} 
  \frac 1 2 &=&-\int_{-\infty}^\infty \frac{dE}{\pi} n_F(E)
  \frac{\Sigma''(E)}{[E+\mu_f-\Sigma'(E)]^2 + [\Sigma''(E)]^2}~. 
\end{eqnarray}

\subsubsection{Kondo temperature\label{section:Kondotemperature}}
On general grounds, the Kondo temperature $T_K$ indicates a crossover between the high temperature weakly coupled and the low temperature strongly coupled regimes. Indeed, the system at temperatures below $T_K$ is characterized by a magnetic confinement of the spin of the edge electronic states which screens the Kondo impurity~\cite{hewson1997}.  
$T_K$ has been shown to be the unique energy scale that characterizes the universal physical properties of single impurity Kondo models at low temperature $T\ll T_K$. We will see later how this scaling property will become extremely useful for analyzing universal properties of the electronic density at very low temperature. But before, we derive here an 
expression of $T_K$. 
Within the mean-field approximation, the Kondo crossover turns to a transition at $T_K$ which corresponds to a continuous vanishing of the $f-c$ effective hybridization: $\Delta(T_K)= 0$. Invoking the self-consistent 
relations~(\ref{eq:mf-general}) and~(\ref{eq:mf-generalbis}), we find (see appendix~\ref{Appendix:subsectionTK}) the following mean-field 
equation for $T_K$: 
\begin{eqnarray}
\mu_f=0~, 
\end{eqnarray}
and
\begin{eqnarray}
\frac{2}{J_K}=P. V. \int_{-\infty}^{+\infty} dE \frac{n_F(E)}{E}
\rho_0(\mu+E)~, 
\label{eq:equationTKgeneral}
\end{eqnarray}
which is equivalent to the Nagaoka-Suhl equation derived in 
Refs.~\onlinecite{nagaoka_resonance} and~\onlinecite{suhl_resonance}. 
Assuming an even free electrons density of state, $\rho_0(-E)=\rho_0(E)$, a general explicit expression of $T_K$ 
was derived in the weak Kondo coupling limit\cite{burdin2000,burdin2009} 
\begin{eqnarray}
T_K&=&F_Ke^{-1/J_K\rho_0(\mu)}~, 
\label{eq:expressionTK-meanfield}
\end{eqnarray}
with
\begin{eqnarray}
F_K&=&\frac{2 e^{\gamma_E}}\pi (D+\mu)
\exp\left[-\int_{-D-\mu}^0
  \frac{dE}{E}\frac{\rho_0(\mu+E)-\rho_0(\mu)}{\rho_0(\mu)}\right]~, 
\label{eq:prefactorTK-meanfield}
\end{eqnarray}
where $\gamma_E\simeq 0.577$ is the Euler-Mascheroni constant, and $D$
denotes the half-bandwidth of $\rho_0$.  
The mean-field expression~(\ref{eq:expressionTK-meanfield}) provides the usual non-analytic exponential 
term characterizing the $J_K-$dependence of the Kondo temperature at small Kondo coupling $J_K\rho_0(\mu)\ll 1$. 
The same exponential dependence also emerges from the mapping to the Bethe Ansatz solvable model 
(see Eq.~\ref{eq:kondo-temp}). 
The prefactor $F_K$ does not depend on $J_K$ and is thus a pure characteristics of the non-interacting electronic system. 
Whatever the specific chemistry of the magnetic impurity is and whatever the microscopic details of its coupling with conduction electrons, the prefactor $F_K$ depends only on the energy structure of the non-interacting conduction electrons. 

We find that $F_K\sim D$ if one assumes a constant density of states and a chemical potential close to the middle of the electronic energy band. The situation may become quantitatively different when the conduction electrons describe the surface modes of a topological insulator. 
Indeed, using the density of states $\rho(E)=|E|/(2\pi v_F^2)$ that characterizes a surface mode without warping, we find 
for $\mu>0$: 
\begin{eqnarray}
  T_K=\frac{2 e^{\gamma_E-1}}{\pi} \frac{\mu^2}{D+\mu} 
\exp{\left( \frac D \mu -\frac{2\pi v_F^2}{J_K\mu}\right) }.  
\end{eqnarray}
This small $J_K$ asymptotic expression becomes singular at the Dirac point $\mu=0$, where the density of states 
vanishes linearly. 
At the Dirac point, one has to start from the equation~(\ref{eq:equationTKgeneral}) for the Kondo temperature, which 
simplifies to:
\begin{eqnarray}
  \label{eq:dirac-point-TK}
  \frac 1 {J_K}&=&\int_0^{D} \frac{dE}{2\pi v_F^2} \tanh
  \left(\frac{E}{2T_K}\right), \\
               &=&\frac{T_K}{\pi v_F^2} \ln\left[\cosh
                 \left(\frac{D}{2T_K}\right)\right],   
\end{eqnarray}
 where $D$ is a bandwidth cutoff. For $J_K<J_K^{c}\equiv\frac{2 \pi v_F^2} D$, the
 Eq.~(\ref{eq:dirac-point-TK}) has no solution, in agreement with the
 results of Ref.~\onlinecite{Withoff1990}. For $J_K>J_K^c$, we find
 that:
 \begin{eqnarray}
   T_K=\frac{D}{2\ln 2} \left(1-\frac{J_K^c}{J_K}\right),  
 \end{eqnarray}
indicating that regular Kondo effect can still be realized at the Dirac point, but with a Kondo temperature 
vanishing linearly with the Kondo interaction when $J_K\to J_K^c$, in agreement with the prediction of
 Ref.~\onlinecite{Withoff1990} for a linear density of states.  

\subsubsection{T-matrix and local electronic density}
The local electronic density is defined as:
\begin{eqnarray}\label{eq:local-density} 
\rho(\mathbf{r})\equiv -\frac{1}{\pi}\mathrm{Im}\left( 
\mathrm{Tr}\left[ G^{cc}(\mathbf{r},\mathbf{r}, i0^{+})\right]\right)
\equiv \rho_0(\mu) + \delta\rho(\mathbf{r})~. 
\end{eqnarray}
From Eqs.~(\ref{eq:gcc-sol}) and~(\ref{eq:gff-sol}) we derive the expression of the T-matrix\cite{affleck2008}, which is defined by the relation : 
\begin{eqnarray} 
  G(\mathbf{r},\mathbf{r'}, i\nu_n)&\equiv&G_0(\mathbf{r}-\mathbf{r'}, i\nu_n)
  + G_0(\mathbf{r}, i\nu_n) T(i\nu_n)
  G_0(-\mathbf{r'}, i\nu_n)~. 
\label{eq:tmatrix-def}
\end{eqnarray}
We find: 
\begin{eqnarray}
 T(i\nu_n)&=&|\Delta|^2G^{ff}(i\nu_n)=|\Delta|^2(i\nu_n+\mu_f
  -\Sigma(i\nu_n))^{-1}~.
\end{eqnarray}
Invoking this expression of the T-matrix and considering that the Kondo self-energy $\Sigma(i\nu_n)$ is diagonal and symmetric in spin components (see Eq.~(\ref{eq:sigma-sol-1})), the local electronic density is then given by : 
\begin{eqnarray}
\delta\rho(\mathbf{r})
=\frac {|\Delta|^2} \beta \sum_{\nu_n}
\frac{\mathrm{Tr}\left[ G_0(\mathbf{r},i\nu_n)
      G_0(\mathbf{-r},i\nu_n)\right] }{ i\nu_n+\mu_f
  -\Sigma(i\nu_n)}~. 
\end{eqnarray}

Assuming that the energy scale which characterizes the Kondo resonance is much smaller than the effective bandwidth of the non-interacting electrons, the T-matrix can be approximated as: 
\begin{eqnarray}
T(i\nu_n)=
\frac{|\Delta|^2}{i(\nu_n + \Gamma
    \mathrm{sign}(\nu_n))}~, 
\end{eqnarray}
where $\Gamma=\pi |\Delta|^2 \rho_0(\mu)$. This expression of the T-matrix is expected to be valid beyond the mean-field approach that we are following here, 
since it is equivalent up to a Fourier transformation to the definitions Eqs. (5.38)--(5.39) on p.112  in
Ref.~\onlinecite{hewson1997}.   

The expression of $\Gamma$ in terms of the microscopic parameters of
the Hamiltonian can be obtained from
Eq.~(\ref{eq:gamma-mf-final-appendix}). For zero temperature, we have
(see App.~\ref{sec:derivation-t-matrix} for a derivation) :
\begin{eqnarray}
\label{eq:kondo-resonance-width} 
  \Gamma(T=0) = 
(D+\mu) \exp\left[ -\frac{1}{J_K\rho_0(\mu)}
-\int_{-D-\mu}^0 \frac {dE}{E}
    \frac{\rho_0(\mu+E)-\rho_0(\mu)}{\rho_0(\mu)}   \right]~. 
\end{eqnarray}  
Comparing this expression with expression~(\ref{eq:expressionTK-meanfield}) for the Kondo temperature, we find the very general (i.e, coupling and band-structure independent) relation 
\begin{eqnarray}
T_K=\frac{2e^{\gamma_E}}{\pi}\Gamma(T=0)~, 
\end{eqnarray}
which connects universally the crossover temperature $T_K$ to the $T=0$ resonance width. 
Invoking the Wilson ratio $R\equiv \chi(T=0)/\gamma=2$, and using the small coupling asymptotic expression 
$\gamma\approx-\frac{1}{\pi}\mathrm{Im}\left( G^{ff}(i0^+)\right)\approx 1/\pi\Gamma$, we find the following 
universal Wilson number within the mean-field approximation: 
\begin{eqnarray}
w\equiv T_K\chi(T=0)=\frac{2e^{\gamma_E}}{\pi^2}\approx 0,36~. 
\end{eqnarray}
This result is in relatively good agreement with Wilson's numerical result $w=0,41$ (see Eq.~(6.75) in 
Ref~\onlinecite{hewson1997}). 
Therefore, the following results will be derived within the mean-field method, but we expect them to be qualitatively and quantitatively valid beyond this approximation.

\subsection{Friedel oscillations without warping}
\label{sec:nowarp}
\begin{figure}[h]
  \centering
  \includegraphics[width=9cm]{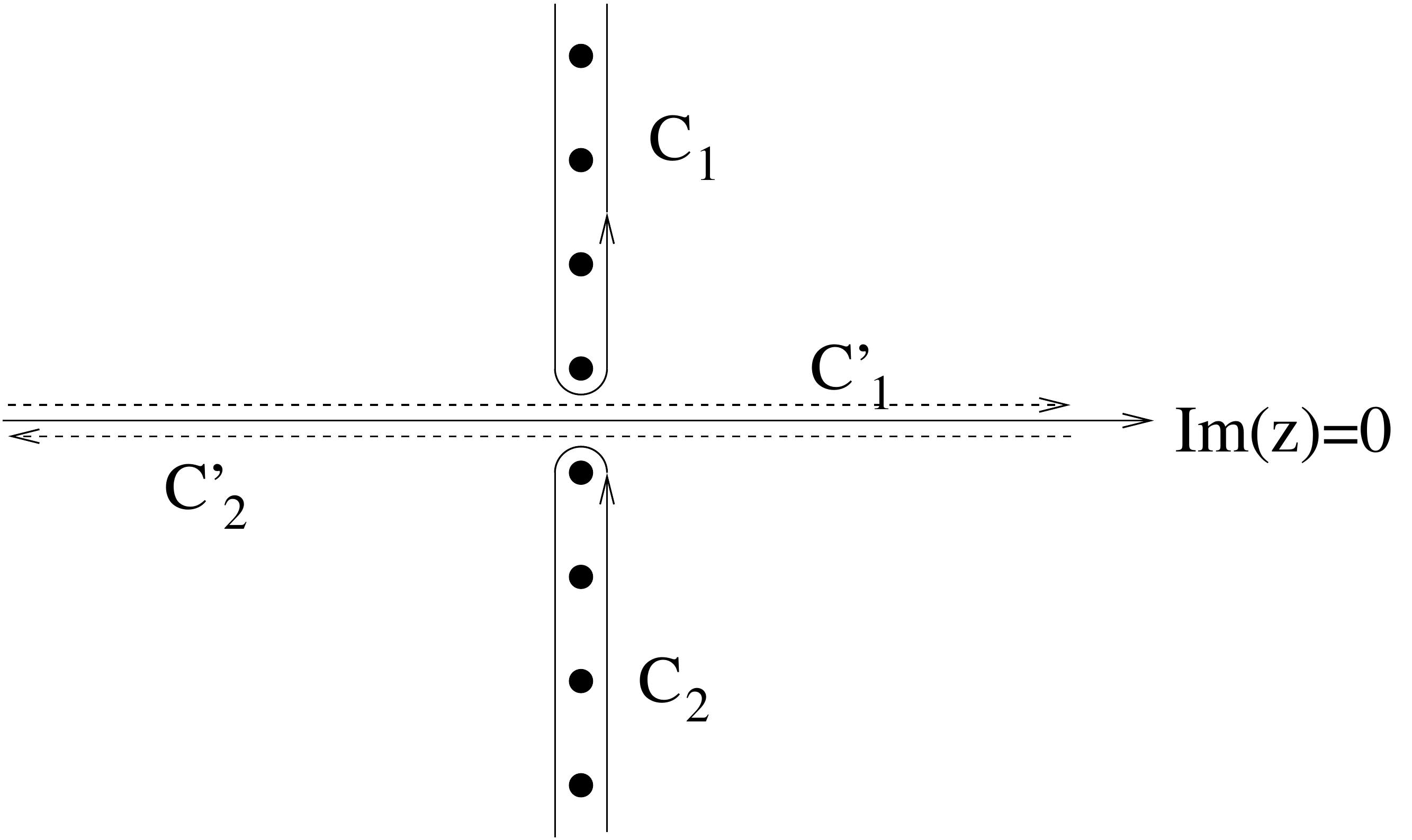} 
  \caption{The integration contours $C_1$ and $C_2$ used in the
    calculation of the sum (\ref{eq:local-density}). These contours
    can be deformed into $C'_1$ and $C'_2$. }
  \label{fig:contour}
\end{figure}

In this section, we study the electronic local density $\rho(\mathbf{r})$ in the absence of warping, i.e., for $\lambda=0$. 
We start from expression~(\ref{eq:local-density}) and we rewrite the sum as a contour integral: 
\begin{eqnarray}
  \label{eq:contour}
  \delta \rho(\mathbf{r}) = -\int_{C_1}
  \frac{dz}{2i\pi} n_F(z)
  \tr\left[G_0(\mathbf{r},z)\frac{|\Delta|^2}{z+i\Gamma}
    G_0(-\mathbf{r},z)\right]   -\int_{C_2}
  \frac{dz}{2i\pi} n_F(z) \tr
  \left[G_0(\mathbf{r},z)\frac{|\Delta|^2}{z-i\Gamma}
    G_0(-\mathbf{r},z)\right],   
\end{eqnarray}
where the contours $C_1$ and $C_2$, depicted by figure~\ref{fig:contour}, encircle the poles of the function
$n_F(z)=(e^{\beta z}+1)^{-1}$ of positive (resp. negative) imaginary
part. The contours can be deformed respectively into $C'_1$ and $C'_2$
allowing to rewrite the integral as: 
\begin{eqnarray}
  \delta \rho(\mathbf{r}) = - |\Delta|^2 \int \frac{dE}{\pi}
  n_F(E)
  \mathrm{Im}\left[\frac{\tr(G_0(\mathbf{r}, E+i0^+)G_0(-\mathbf{r},E+i0^+))}{E+i\Gamma} \right]~, 
\end{eqnarray}
where $E$ covers the real axis and $G_0(\mathbf{r},E+i0^+)$ will be obtained in the following from analytic continuation. 
Indeed, in the absence of warping, an analytic
expression of the Green's function is available from the non-interacting 
Hamiltonian~(\ref{eq:warping}). After Fourier transform of 
$(i\nu_n-{\cal H}^0(\mathbf{k}))^{-1}$, we find : 
 \begin{eqnarray}\label{eq:nowarp-gf} 
   G_0(\mathbf{r},i\nu_n)&=&\frac{i\nu_n +\mu}{2\pi v_F^2} \left[
     K_0 \left(\frac {|\nu_n|-i\mu~\mathrm{sign}(\nu_n)}{v_F} r \right) +\mathrm{sign}(\nu_n)
     \mathbf{\hat{z}}\cdot(\boldsymbol{\sigma}\times \mathbf{\hat{r}})
     K_1\left(\frac {|\nu_n|-i\mu~\mathrm{sign}(\nu_n)}{v_F} r\right)  \right]~, 
 \end{eqnarray}
and 
\begin{eqnarray}
  \tr\left[G_0(\mathbf{r},i\nu_n) G_0(-\mathbf{r},i\nu_n) \right] =
\frac{(i\nu_n+\mu)^2}{2\pi^2 v_F^4}
\left[K_0\left(\frac{|\nu_n|-i\mu~\sign(\nu_n)} {v_F} r\right)^2 -
  K_1\left(\frac{|\nu_n|-i\mu~\sign(\nu_n)} {v_F} r\right)^2 \right]~, 
\end{eqnarray}
where $K_0$ and $K_1$ are respectively modified Bessel functions the
second kind of order zero and one,\cite{abramowitz_math_functions}  
and $r\equiv\vert\mathbf{r}\vert$. 
Hereafter we will replace this expression by its analytic continuation in the upper half complex plane, 
$i\nu_n\to E+i0^+$, with $\nu_n>0$. 
Using Eq. (9.6.4) in
Ref.~\onlinecite{abramowitz_math_functions}, we have:
\begin{eqnarray}
  K_\nu\left(\frac{0^+ -i(E+\mu)}{v_F} r\right) = i\frac \pi 2
  e^{i\frac \pi 2 \nu} H_{\nu}^{(1)} \left(\frac{(E+\mu +i 0^+)
      }{v_F}r\right)~, 
\end{eqnarray}
where $H_\nu^{(1)}$ denotes the Hankel function.\cite{abramowitz_math_functions} 
We thus find the expression:
\begin{eqnarray}
\delta \rho(\mathbf{r})= \frac{|\Delta|^2}{8 \pi v_F^4 }  \int
 dE~(E+\mu)^2~n_F(E)~
  \mathrm{Im}\left[\frac{(H_0^{(1)})^2\left(\frac{(E+\mu) r}{v_F}\right) +(H_1^{(1)})^2\left(\frac{(E+\mu) r}{v_F}\right)}{E+i \Gamma} \right]~. 
\end{eqnarray}

For long distances, $r\gg R_F=v_F/\mu$, we can use the approximation
from Ref.~\onlinecite{abramowitz_math_functions}:
\begin{eqnarray}
  \label{eq:approx-hankel}
  (H_0^{(1)})^2(z)+(H_1^{(1)})^2(z) = \frac{-2}{\pi z^2}
    e^{i(2z-\frac \pi 2)} + O(1/z^3), 
\end{eqnarray}
giving: 
\begin{eqnarray}\label{eq:approx-nowarp} 
\delta \rho(\mathbf{r}) \simeq 
\frac{\Gamma}{2\pi^2 \mu r^2} 
\int_{-\infty}^\infty \frac{d\zeta}{1+\exp{\left( \frac{E}T\right) }}
  \mathrm{Re}\left[\frac{\exp{\left( 2i \left( E+\mu\right)r/v_F\right) }}{
E+i \Gamma} \right]~, 
\end{eqnarray}
where we have used the relation $\vert\Delta\vert^2=\Gamma/\pi\rho_0(\mu)=2\Gamma v_F^2/\mu$.

At zero temperature, with Eq.(\ref{eq:approx-nowarp}) 
we find the following expression for the local density:
\begin{eqnarray}\label{eq:nowarp-gs} 
  \delta \rho(\mathbf{r})&=&-\frac{\Gamma}{2\pi^2 \mu r^2}
  e^{\frac{2\Gamma r}{v_F}} E_1\left(\frac {2\Gamma r} {v_F} \right) \cos
  \left(\frac{2\mu r}{v_F}\right)~, 
\end{eqnarray}
where $E_1(u)$ is the exponential
integral\cite{abramowitz_math_functions}. For long distance $r \gg
v_F/\Gamma$ , since $e^u
E_1(u) \sim 1/u$, we obtain  
\begin{equation}
  \label{eq:nowarp-gs-far-imp}
  \delta\rho(\mathbf{r}) \sim  -\frac{v_F}{4\pi^2 \mu r^3} \cos
  \left(\frac{2\mu r}{v_F}\right).   
\end{equation}
Remarkably, the amplitude of the Friedel oscillations become
independent of $\Gamma$, i.e., independent of $T_K$, 
for distances longer than the Kondo lengthscale $R_K\equiv v_F/\Gamma$. 
This can be understood by noting that for distances
larger than $R_K$, the impurity appears as a potential
scatterer at unitarity.\cite{nozieres_shift} For short distances $r
\ll v_F/(2\mu), R_K$, we
find
\begin{equation}
  \label{eq:nowarp-gs-close}
  \delta \rho(\mathbf{r}) = \frac{\Gamma}{2\pi^2 \mu r^2} \ln
  \left(\frac{2 \Gamma e^\gamma r}{v_F}\right),  
\end{equation}
so that the amplitude of oscillations inside the Kondo cloud depends explicitly on $\Gamma$. 
This indicates that the Friedel oscillations at distances shorter than
the Kondo lengthscale $R_K$ reflect the internal structure of the Kondo
screening cloud.\cite{affleck2008}

\begin{figure}
\centering
\begin{minipage}[c]{\linewidth}
\begin{minipage}[r]{0.5\linewidth}
\includegraphics[width=9cm, origin=tr]{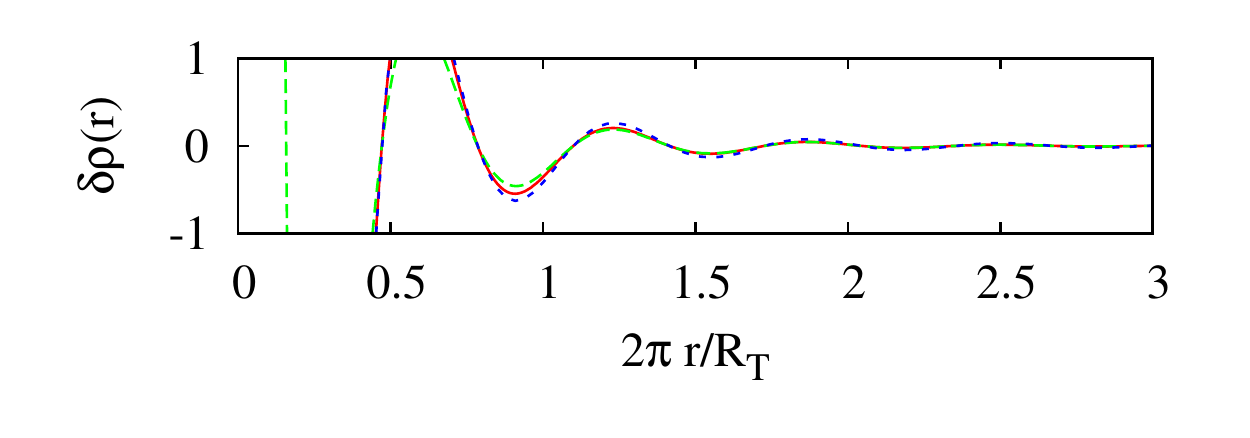}
\end{minipage}
\begin{minipage}[c]{0.1\linewidth}
a)
\end{minipage}\\
\begin{minipage}[r]{0.5\linewidth}
\vspace*{-.5cm}\includegraphics[width=9.25cm, origin=tr]{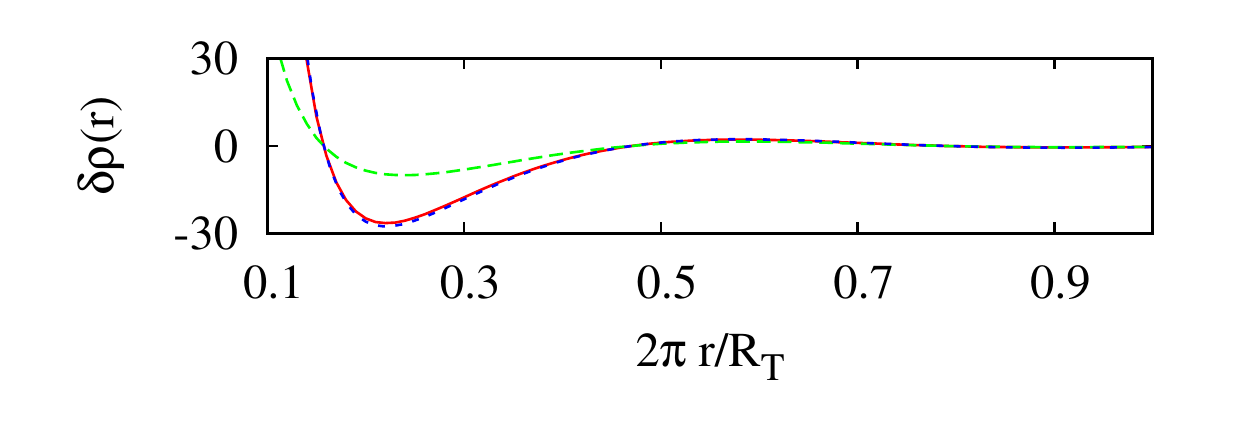}\hspace*{0.5cm}
\end{minipage}
\begin{minipage}[c]{0.1\linewidth}
\vspace*{-.5cm}b)
\end{minipage}
\end{minipage}
\caption{(Color online) Variation of electronic density computed at finite temperature using the exact 
expression Eq.~(\ref{eq:nowarp-temp}) [red solid line], the long distance coherent approximation 
Eq.~(\ref{eq:nowarp-gs-far-imp}) [blue doted line] which is appropriate for $R_K<r<R_T$, and the very long distance 
regime $r> R_T$ approximated by Eq.~(\ref{eq:thermalreductionnowarping}) [green dashed line]. 
a) Overview along few Friedel oscillations, b) Focus at relatively short distance. 
Here, we chose $R_T=50R_K$ and $\pi R_F=5R_K$. }
\label{fig:oscillations}
\end{figure}

\begin{figure}
\centering
\includegraphics[width=9cm]{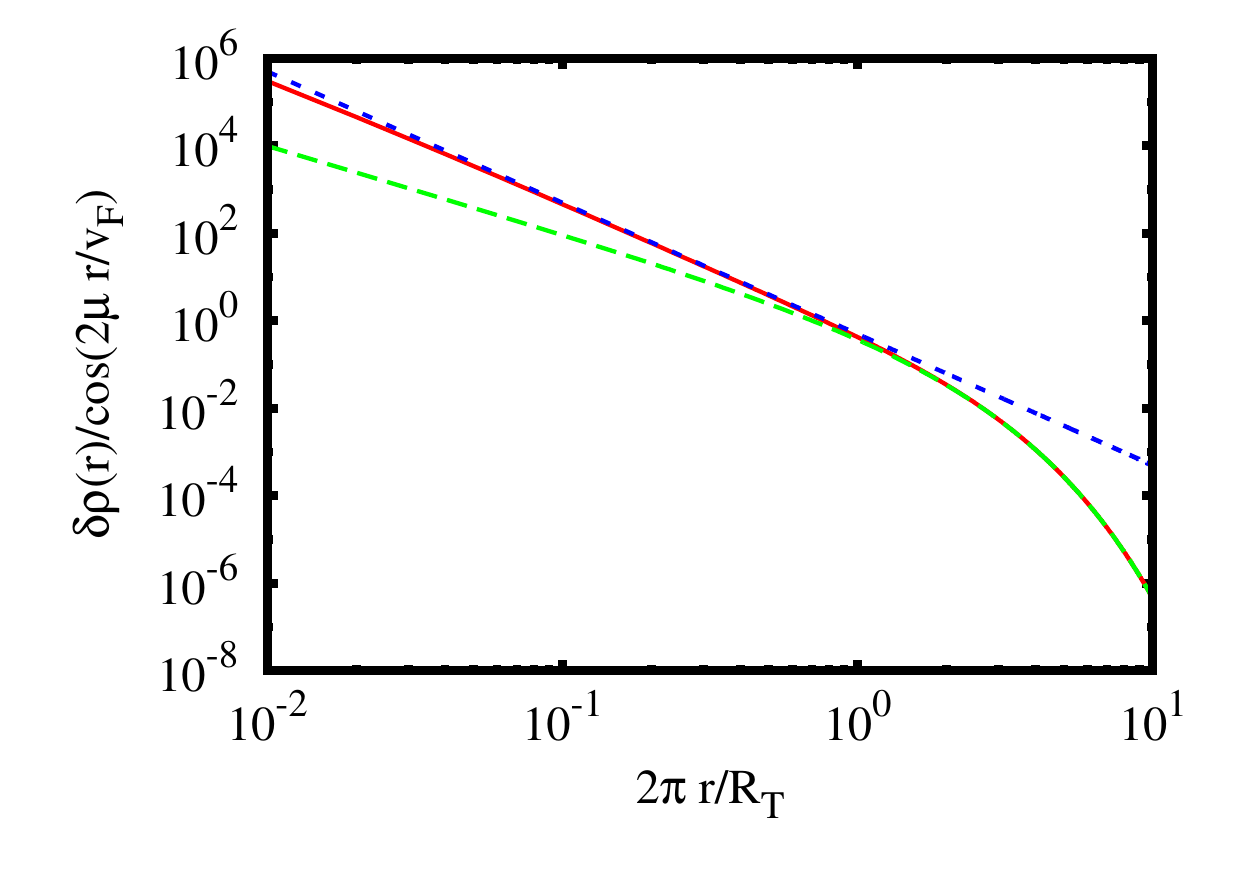} 
\caption{(Color online) Plot of the envelope of Friedel oscillations, $\delta\rho ({\bf r })/\cos(2\mu r/v_F)$, 
computed at finite temperature using the exact 
expression Eq.~(\ref{eq:nowarp-temp}) [red solid line], the long distance coherent approximation 
Eq.~(\ref{eq:nowarp-gs-far-imp}) [blue dotted line] which is appropriate for $R_K<r<R_T$, and the very long distance 
regime $r> R_T$ approximated by Eq.~(\ref{eq:thermalreductionnowarping}) [green dashed line]. 
Here, we chose $R_T=50R_K$ and $\pi R_F=5R_K$. }
\label{fig:envelopes}
\end{figure}

At finite temperature, Eq.(\ref{eq:approx-nowarp})  yields:
\begin{eqnarray}\label{eq:nowarp-temp}   
  \delta \rho(\mathbf{r})&=& -\frac{\Gamma}{2\pi^2 \mu r^2}
  \frac{e^{-\frac{2\pi r}{\beta v_F}}}{\frac 1 2 + \frac{\beta
      \Gamma}{2\pi}} {}_2F_1\left(1,\frac 1 2 + \frac{\beta
      \Gamma}{2\pi}; \frac 3 2 + \frac{\beta
      \Gamma}{2\pi}; e^{-\frac{4\pi r}{\beta v_F}} \right) \cos
  \left(\frac{2\mu r}{v_F}\right)~, 
\end{eqnarray} 
where ${}_2F_1$ is the Gauss hypergeometric
function\cite{abramowitz_math_functions}. For long distances, and
$T\ll T_K$ the Friedel oscillations decay exponentially over the thermal length
$R_T\equiv v_F/T$, 
\begin{equation}
\label{eq:thermalreductionnowarping}
  \delta \rho(\mathbf{r}) \sim -\frac{e^{\frac{-2\pi r}{\beta v_F}}}{ \pi 
    \beta \mu r^2} \cos\left(\frac{2\mu r}{v_F}\right),  
\end{equation}
 in agreement with Ref.~\onlinecite{tran2010}.  The behavior is
 represented on Fig.~\ref{fig:oscillations} which depicts the Friedel oscillations, and 
on Fig.~\ref{fig:envelopes} which depicts the envelope of these oscillations. 
It appears clearly that the Friedel oscillations have a period $\pi R_F$, and an envelope 
decreasing like $1/r^3$ when $R_K<r<R_T$ as approximated by Eq.~(\ref{eq:nowarp-gs-far-imp}). 
Their amplitude is exponentially reduced at longer distances $r> R_T$ where Eq.~(\ref{eq:thermalreductionnowarping}) 
provides a good approximation.
\begin{figure}[h]
  \centering
  \includegraphics[width=9cm]{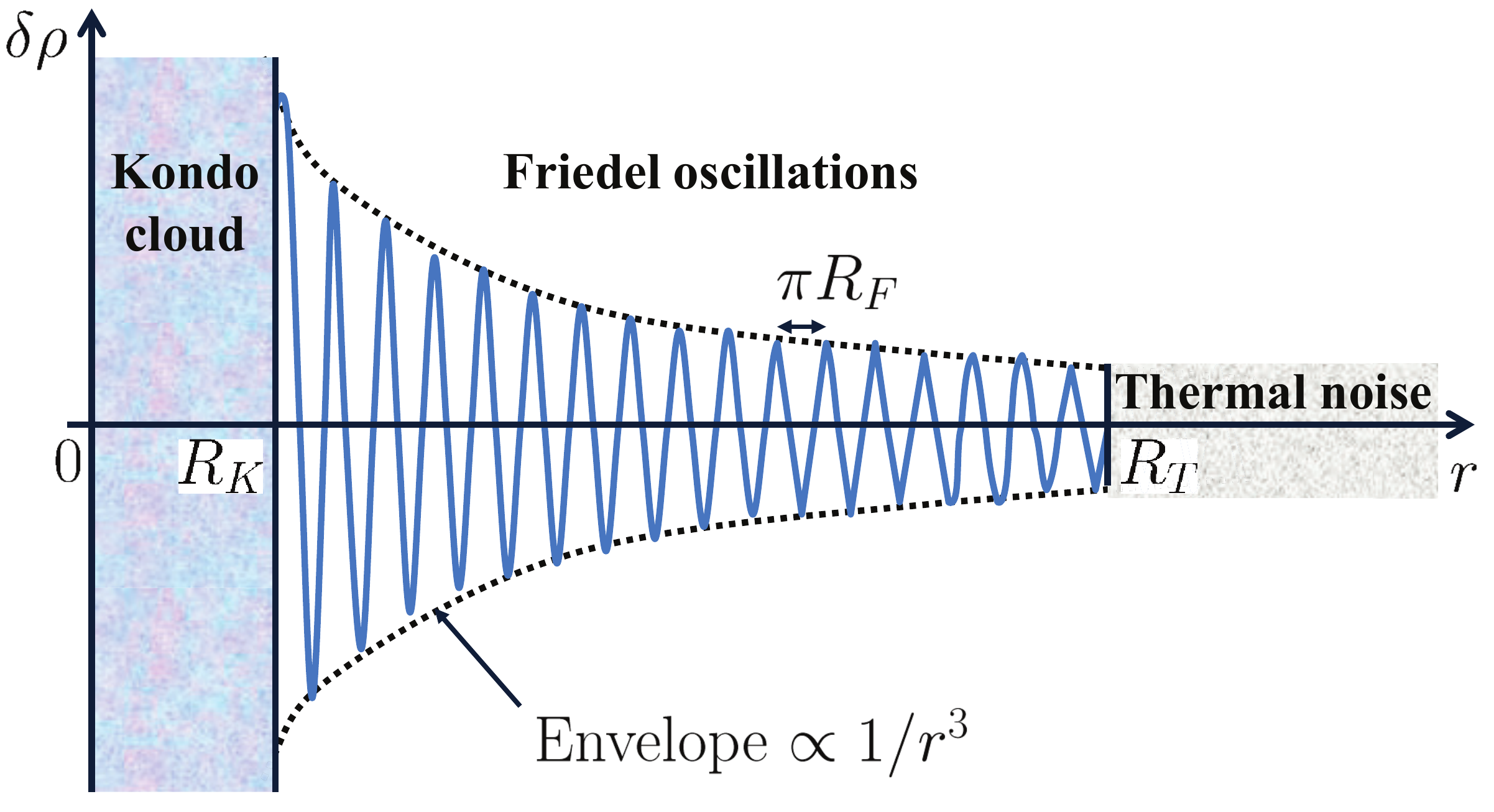} 
  \caption{Schematic description of the Friedel oscillations that can be observed in the density variations around the impurity when $T<T_K$. For the rotationally invariant case $\lambda=0$ the envelope decays like $1/r^3$. For the warping 
term $\lambda\neq 0$ the decay is slower (like $1/r^2$) in some directions. }
  \label{fig:schemaFriedel}
\end{figure}
We see from the previous analysis 
that there are three relevant lengthscales as depicted schematically by figure~\ref{fig:schemaFriedel}. The first one, $R_F = v_F/\mu$, is proportional to  
the Fermi wavelength. For lengthscales much
larger than $R_F$, the simplification~(\ref{eq:approx-hankel})
is justified, and $R_F$ then simply gives the pseudo-period of the Friedel
oscillations. The second lengthscale is the Kondo screening length
$R_K$, which is the size of the Kondo cloud; due to the temperature dependence of 
the Kondo resonance width, $\Gamma(T)$, we have $R_K(T\gtrsim T_K)=\infty$ and 
$R_K( T\ll T_K)\propto v_F/T_K$. 
Above that
lengthscale, which requires at least $T<T_K$, the Friedel oscillations become identical to those of 
a resonant non-magnetic impurity. The third important lengthscale is
the thermal length $R_T$. Beyond that length, the Friedel
oscillations decay exponentially, while below $R_T$ they are
unchanged from the zero temperature case. In order to observe the Friedel
oscillations characteristic of the Kondo screening cloud, we must have
$T\ll T_K$ so that the Kondo screening length is much shorter than the
thermal length. In a renormalization group picture, the temperature 
is the natural infrared cutoff for the renormalization group flow,
and the constraint $T\ll T_K$ is simply a requirement that the strong
coupling scale is reached before the thermal cutoff.

\subsection{Friedel oscillations with warping}
\label{sec:warp}
We now turn to 
 the electronic local density $\rho(\mathbf{r})$ in the presence of warping. 
For $\lambda\neq 0$ we don't have anymore an expression in
closed form of the Green's function. Instead, we use the
approximation~(\ref{eq:green-asymp-warp}) 
derived in App.\ref{app:asymp}~: 
 \begin{eqnarray}\label{eq:warp-approx} 
   G_0(\mathbf{r},i\nu_n)=G_0(r, \theta,i\nu_n)&=&\frac{i\nu_n +\mu}{2\pi v_F^2} \left[
     K_0 (\zeta_n) +\mathrm{sign}(\nu_n)
     \mathbf{\hat{z}}\cdot(\boldsymbol{\sigma}\times \mathbf{\hat{r}})
     K_1(\zeta_n)  \right. \\ && \left. 
+ \lambda \sigma^z\cos (3\theta ) 
     (i\nu_n+\mu)^2  \mathrm{sign}(\nu_n)
     K_3(\zeta_n)/v_F^3  \right] + O(\lambda^2)~, 
\end{eqnarray}
with
\begin{eqnarray}
\zeta_n &\equiv& \frac {|\nu_n|-i\mu~\mathrm{sign}(\nu_n)}{v_F}r~. 
\end{eqnarray}
We obtain:
\begin{eqnarray}
\tr\left[G(r,\theta,i\nu_n) G(r,\theta+\pi,i\nu_n) \right] =
\frac{(i\nu_n+\mu)^2}{2\pi^2 v_F^4}
\left[K_0\left(\frac{|\nu_n|-i\mu~\sign(\nu_n)} {v_F} r\right)^2 -
  K_1\left(\frac{|\nu_n|-i\mu~\sign(\nu_n)} {v_F} r\right)^2 
\right. \nonumber \\ -\left. 
  \left(\frac \lambda {v_F} \cos 3\theta\right)^2 \left(\frac{i\nu_n
      +\mu}{v_F}\right)^4   K_3\left(\frac{|\nu_n|-i\mu~\sign(\nu_n)} {v_F} r\right)^2 \right]~.  
\end{eqnarray}
Since this expression is only approximate, we cannot use contour
integration techniques to obtain the sum~(\ref{eq:local-density}). 
Indeed, attempting to use contour integration for $T>0$ yields a divergent
integral. Nevertheless, for zero temperature, we can still 
change the sum (\ref{eq:local-density}) into an integral as the
exponential decay of the modified Bessel function ensures the
convergence of the sums.  This leads to the zero temperature result
valid for $r\gg R_K=v_F/\Gamma$ : 
    \begin{eqnarray}\label{eq:warp-gs}      
      \delta \rho(\mathbf{r})&=&\delta \rho(\mathbf{r})|_{\lambda=0}
    -\lambda^2 \frac{\mu^4  \cos^2 (3\theta) }{4 \pi^2 v_F^6 r^2}
\cos \left (2\frac{\mu r}{v_F}\right)~. 
    \end{eqnarray}
For positive temperature, we have to compute the
sum~(\ref{eq:local-density}) numerically. This will be done in the next section using realistic values for the 
model parameters. 
Here we rather discuss general new features that emerge from this warping term. 
First, we expect finite temperature corrections to be relevant only in the crossover temperature regime 
around $T_K$. Indeed, similarly to what we found without warping, the thermal length $R_T=v_F/T$ provides a 
cut-off below which Friedel oscillations are identical to the ones predicted for $T=0$, and above which they are 
muffled by thermal fluctuations. 

Furthermore, one important feature here in this expression is the $1/r^2$ decay of the envelope: this decay is identical to the one of a two-dimensional normal metal, and it dominates the $1/r^3$ contribution from the non-warping. Because of the 
$\cos^2 3\theta$ factor, the Friedel oscillations in the directions $\theta=\frac{\pi}{6}(2n+1)$ have the contribution from warping switched off and contain only  the $1/r^3$ contribution, whilst in other directions the warping contribution is observable and dominates on the longer lengthscales due to its slower $1/r^2$ decay.  

Also, $R_K=v_F/\Gamma$ characterizing the size of the Kondo screening cloud, we expect the Friedel oscillations to be observable only at distance larger than this Kondo length. Within a renormalization group picture, the density oscillations are thus supposed to be measured at a sufficiently large distance from the Kondo impurity such that the system 
is correctly described by its strong coupling fixed point, i.e., the Kondo spin is totally screened. Nevertheless, comparing 
Eq.~(\ref{eq:nowarp-gs-far-imp}) and Eq.~(\ref{eq:warp-gs}) we find that the warping correction becomes relevant 
only for distances larger than a new characteristic length,
$R_W=\frac{v_F^7}{\lambda^2\mu^5}$.  
Introducing the density of surface states $n_S$ and invoking the
density of states $\rho_0(E)=\vert E\vert/(2\pi v_F^2)$ in the
vicinity of the Dirac point, we have $n_S=\mu^2/(4\pi v_F^2)=1/(4\pi
R_F^2)$, so that $R_W=(v_F/\lambda)^2(4\pi n_S)^{-5/2}$. 
A crossover density emerges, $n_S^\star\equiv\frac 1 {4\pi}
\left(\frac{v_F^2}{\lambda^2 R_K}\right)^{2/5}$, that distinguishes two different cases: 
for $n_S>n_S^\star$, we find $R_W<R_K$ and the Friedel oscillations which are observed for 
$R_K<r<R_T$ are characterized by the warping term with switch on and off directions and an $1/r^3$ envelope. 
But, closer to the Dirac point, i.e., for density $n_S<n_S^\star$, Friedel oscillations are characterized by two regimes: 
at shortest distances $R_K<r<R_W$ the isotropic term with $1/r^2$ envelope dominates, whilst  
the warping correction dominates at larger distances $R_W<r<R_T$. Also, a new temperature scale emerges
 in the lowest density case $n_S<n_S^\star$: the warping temperature $T_W\equiv v_F/R_W<T_K$ indicating the 
crossover temperature below which warping effects appear.

\section{Discussion}
\subsection{Experimental observability of the density oscillations}
Here we analyze the experimental observability of the density oscillation effects, with or without warping effects, that were discussed on general grounds in the previous section. 
The idea is to compute the density variation $\delta\rho ({\bf r)}$ around a Kondo impurity using 
realistic values of parameters that correspond to topologically insulating compounds for which surface 
states have been observed or predicted. We consider more specifically two compounds: $\mathrm{Bi_2Se_3}$ for which  Ref.~\onlinecite{kuroda2010} gives $v_F=3.55$~eV\AA~and $\lambda=128$ eV\AA$^3$, and $\mathrm{Bi_2Te_3}$ 
with values $v_F=2.55$ eV\AA~and $\lambda=250$ eV\AA$^3$~given by
Refs.~\onlinecite{chen2009} and~\onlinecite{an2012}. 
We are still left with two tunable parameters: the Kondo temperature $T_K=v_F/R_K$, 
and the density of surface states $n_S=\mu^2/(4\pi v_F^2)=1/(4\pi R_F^2)$.  
First, we remark that the Kondo temperature depends on various microscopic parameters including the chemistry of the magnetic impurity and the density $n_S$. Furthermore the well know exponential $J_K-$dependence of $T_K$ 
(see section~\ref{section:Kondotemperature}) 
makes this temperature scale very sensitive to variations of these microscopic parameters. Therefore, refereeing from 
the orders of magnitude that are usually measured in Kondo compounds we consider here three different characteristic values: $T_K=1000$~K (big), $T_K=100$~K (medium), and $T_K=10$~K (relatively small). 
For the density of surface states $n_S$, we consider three values for each compound: 
the crossover density $n_S^\star= \frac 1 {4\pi}
\left(\frac{v_F^2}{\lambda^2 R_K}\right)^{2/5}$, a smaller density $n_S=n_S^\star/10$, and a larger density 
$n_S=10n_S^\star$. 
The sets of relevant parameters that we consider are summarized in table~\ref{Table:Bi2Se3} for $\mathrm{Bi_2Se_3}$, 
and in table~\ref{Table:Bi2Te3} for $\mathrm{Bi_2Te_3}$. 
\begin{table}
\begin{tabular}{|l|l|l|l|}
\hline
~~&$T_K=10$~K&$T_K=100$~K&$T_K=1000$~K\\
~~&$R_K=410$~nm&$R_K=41$~nm&$R_K=4,1$~nm\\
\hline
$n_S\approx n_S^\star/10$&$n_S=1,6.10^3$~$\mu$m$^{-2}$&$n_S=4.10^3$~$\mu$m$^{-2}$&
$n_S=10^4$~$\mu$m$^{-2}$\\
~~&$\mu=50$~meV&$\mu=80$~meV&$\mu=130$~meV\\
~~&$R_F=7,0$~nm&$R_F=4,4$~nm&$R_F=2,8$~nm\\
~~&$R_W=130$~$\mu$m&$R_W=13$~$\mu$m&$R_W=1,3$~$\mu$m\\
~~&$T_W=3.10^{-5}$~K&$T_W=3.10^{-4}$~K&$T_W=3.10^{-3}$~K\\
\hline
$n_S\approx n_S^\star$&$n_S=1,6.10^4$~$\mu$m$^{-2}$&$n_S=4.10^4$~$\mu$m$^{-2}$&
$n_S=10^5$~$\mu$m$^{-2}$\\
~~&$\mu=160$~meV&$\mu=250$~meV&$\mu=400$~meV\\
~~&$R_F=2,2$~nm&$R_F=1,4$~nm&$R_F=8,8$~\AA\\
~~&$R_W=410$~nm&$R_W=41$~nm&$R_W=4,1$~nm\\
\hline
$n_S\approx 10n_S^\star$&$n_S=1,6.10^5$~$\mu$m$^{-2}$&$n_S=4.10^5$~$\mu$m$^{-2}$&
$n_S=10^6$~$\mu$m$^{-2}$\\
~~&$\mu=500$~meV&$\mu=800$~meV&$\mu=1,3$eV\\
~~&$R_F=7,0$~\AA&$R_F=4,4$~\AA&$R_F=2,8$~\AA\\
~~&$R_W=1,3$~nm&$R_W=1,3$~\AA&$R_W=0,13$~\AA\\
\hline
\end{tabular}
\caption{Model parameters for $\mathrm{Bi_2Se_3}$, with 
$v_F=3.55$~eV\AA~and $\lambda=128$ eV\AA$^3$.  }
\label{Table:Bi2Se3}
\end{table}
\begin{table}
\begin{tabular}{|l|l|l|l|}
\hline
~~&$T_K=10$~K&$T_K=100$~K&$T_K=1000$~K\\
~~&$R_K=300$~nm&$R_K=30$~nm&$R_K=3$~nm\\
\hline
$n_S\approx n_S^\star/10$&$n_S=8,3.10^2$~$\mu$m$^{-2}$&$n_S=2,1.10^3$~$\mu$m$^{-2}$&
$n_S=5,2.10^3$~$\mu$m$^{-2}$\\
~~&$\mu=26$~meV&$\mu=41$~meV&$\mu=65$~meV\\
~~&$R_F=9,8$~nm&$R_F=6,2$~nm&$R_F=3,9$~nm\\
~~&$R_W=94$~$\mu$m&$R_W=9,4$~$\mu$m&$R_W=940$~nm\\
~~&$T_W=3.10^{-5}$~K&$T_W=3.10^{-4}$~K&$T_W=3.10^{-3}$~K\\
\hline
$n_S\approx n_S^\star$&$n_S=8,3.10^3$~$\mu$m$^{-2}$&$n_S=2,1.10^4$~$\mu$m$^{-2}$&
$n_S=5,2.10^4$~$\mu$m$^{-2}$\\
~~&$\mu=82$~meV&$\mu=130$~meV&$\mu=210$~meV\\
~~&$R_F=3,1$~nm&$R_F=2,0$~nm&$R_F=1,2$~nm\\
~~&$R_W=300$~nm&$R_W=30$~nm&$R_W=3$~nm\\
\hline
$n_S\approx 10n_S^\star$&$n_S=8,3.10^4$~$\mu$m$^{-2}$&$n_S=2,1.10^5$~$\mu$m$^{-2}$&
$n_S=5,2.10^5$~$\mu$m$^{-2}$\\
~~&$\mu=260$~meV&$\mu=410$~meV&$\mu=650$~meV\\
~~&$R_F=9,8$~\AA&$R_F=6,2$~\AA&$R_F=3,9$~\AA\\
~~&$R_W=9,3$~\AA&$R_W=0,93$~\AA&$R_W=0,093$~\AA\\
\hline
\end{tabular}
\caption{Model parameters for $\mathrm{Bi_2Te_3}$, with 
$v_F=2.55$ eV\AA~and $\lambda=250$ eV\AA$^3$.  }
\label{Table:Bi2Te3}
\end{table}

\begin{figure}
\centering
\includegraphics[width=8cm,angle=0]{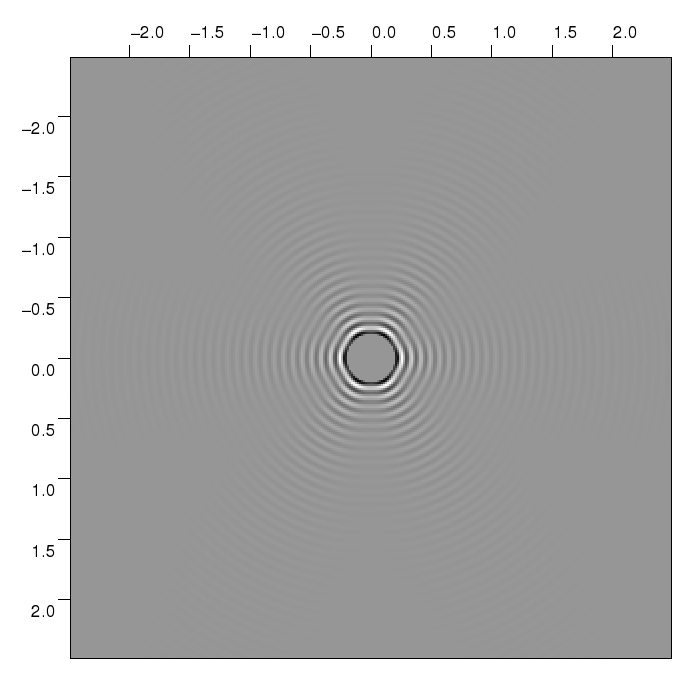} 
\includegraphics[width=8cm,angle=0]{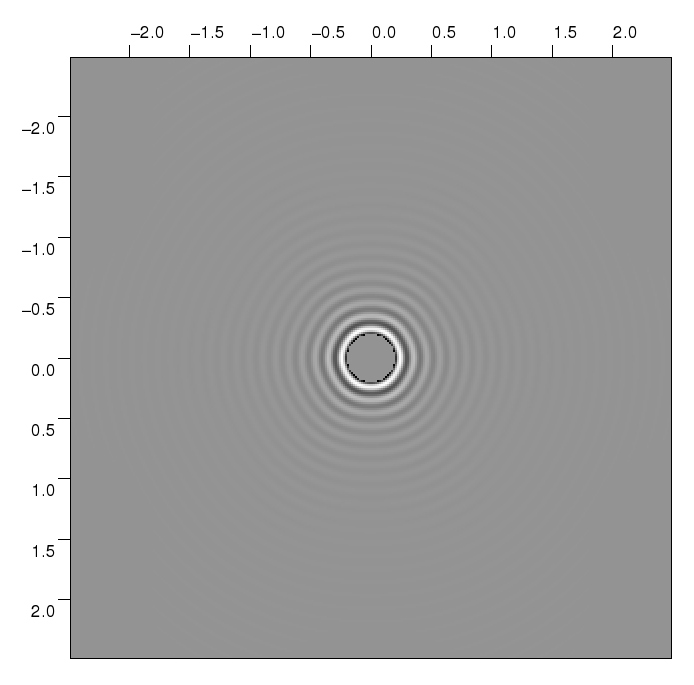} 
\includegraphics[width=8cm,angle=0]{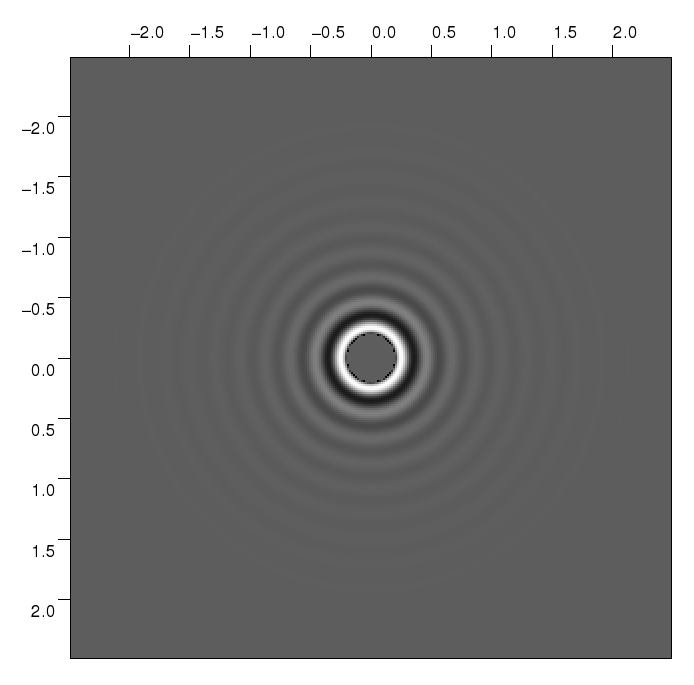} 
\includegraphics[width=8cm,angle=0]{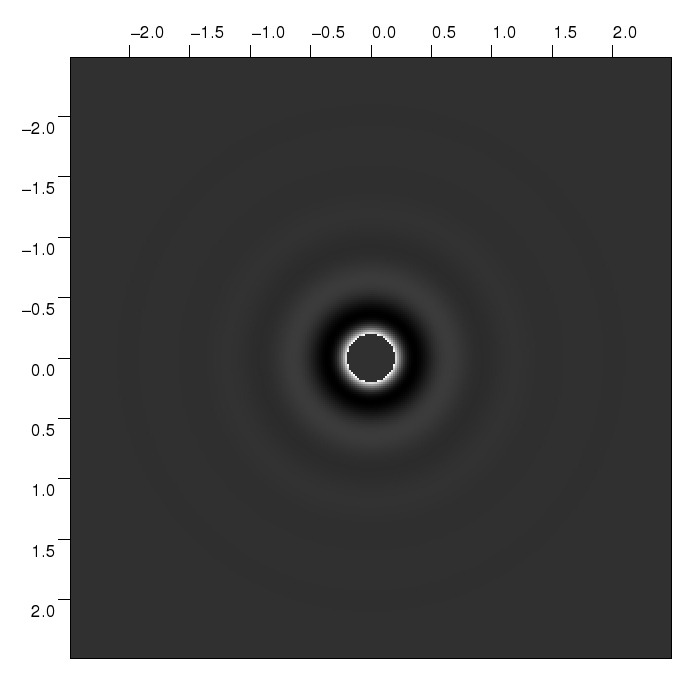} 
\caption{Intensity plot of the Friedel oscillations around a screened magnetic impurity at the edge of 
$\mathrm{Bi_2Se_3}$. Model parameters $v_F=3.55$~eV\AA~and $\lambda=128$ eV\AA$^3$. 
We chose $T_K=100K$ and $T=10K$. From top left to bottom right, $\mu=350$~meV, $250$~meV, $130$~meV, $50$~meV. Unit length: $R_K$. Short distances, $r<0.2 R_K$, are not represented. 
Intensity in arbitrary units is represented by the darkness of the plots. 
}
  \label{fig:intensityBi2Se3}
\end{figure}

\begin{figure}
\centering
\includegraphics[width=8cm,angle=0]{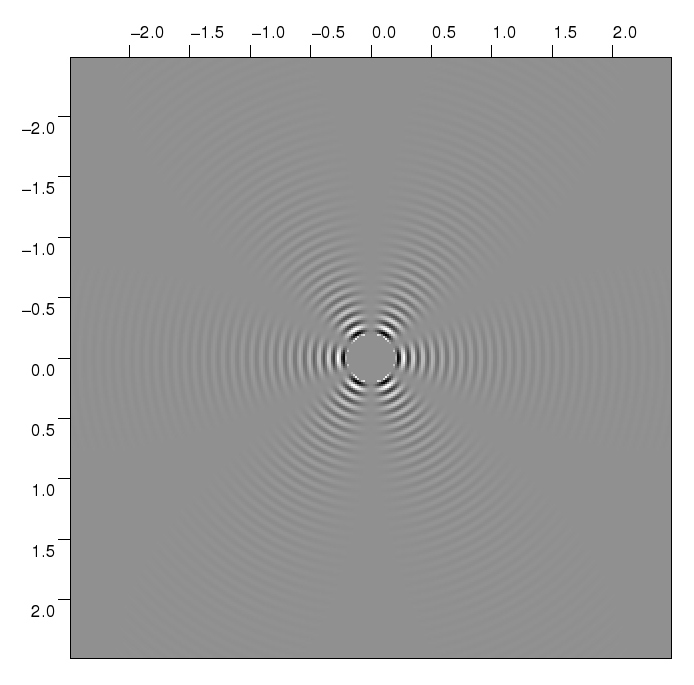} 
\includegraphics[width=8cm,angle=0]{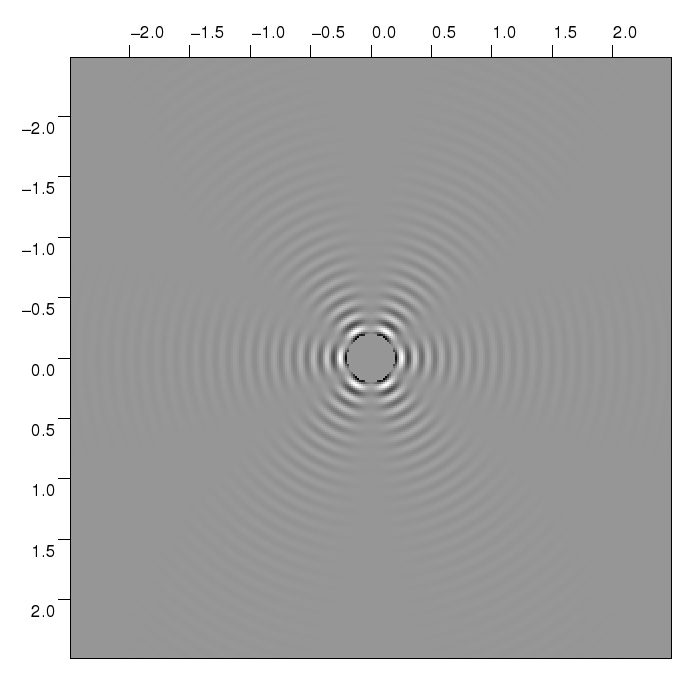} 
\includegraphics[width=8cm,angle=0]{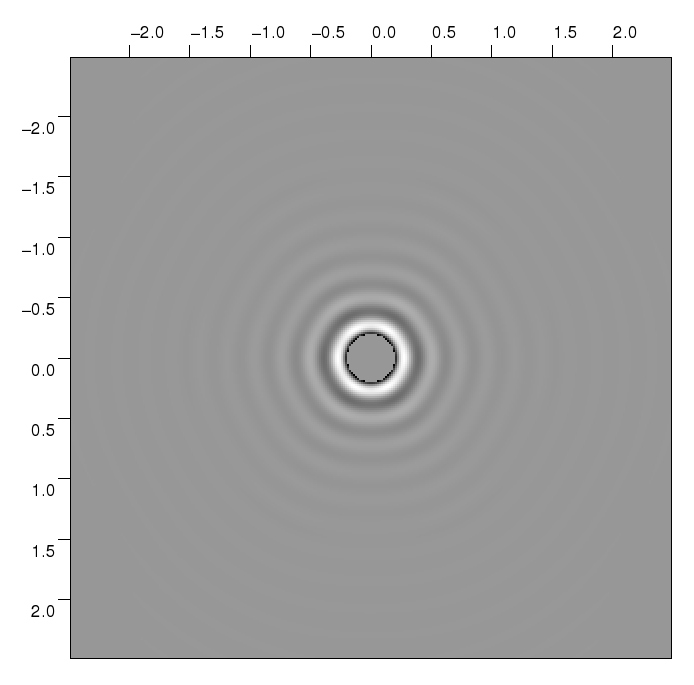} 
\caption{Intensity plot of the Friedel oscillations around a screened magnetic impurity at the edge of 
$\mathrm{Bi_2Te_3}$. Model parameters $v_F=2.55$~eV\AA~and $\lambda=250$ eV\AA$^3$. 
We chose $T_K=100K$ and $T=10K$. From left to right, $\mu=340$~meV, $250$~meV, $120$~meV. 
Unit length: $R_K$. Short distances, $r<0.2 R_K$, are not represented. 
Intensity in arbitrary units is represented by the darkness of the plots. 
}
  \label{fig:intensityBi2Te3}
\end{figure}

The results are represented on Figs.~\ref{fig:intensityBi2Se3} and~\ref{fig:intensityBi2Te3} 
for $\mathrm{Bi_2Se_3}$ and $\mathrm{Bi_2Te_3}$ respectively. 
For these plots, we fixed arbitrarily $T_K=100$~K and we chose
realistic relevant values of chemical potential $\mu$, that can be
controlled experimentally by doping with Sn\cite{chen2009} or Mg\cite{kuroda2010}. In $\mathrm{Bi_2Se_3}$ compounds, experimental values of $\mu$ indicated in Ref.~\onlinecite{kuroda2010} are tuned from $350$~meV down to $0$~eV. Therefore, the four values that we considered for the plots of Fig.~\ref{fig:intensityBi2Se3} were chosen invoking table~\ref{Table:Bi2Se3} 
in order to illustrate the observability of the various cases: with dominant warping term  ($\mu=350$~meV$>\mu^\star$), 
with similar warping and isotropic terms ($\mu=\mu^\star=250$~meV), and with negligible warping term ($\mu=130$~meV, and $50$~meV). Figure~\ref{fig:intensityBi2Se3} clearly shows the Friedel oscillations  
with six-fold rotation symmetry when $\mu>\mu^\star$, or with full rotation symmetry when $\mu<\mu^\star$. 
The choice of $T_K=100$~K for this plot is arbitrary and experimental values of the
Kondo temperature can be significantly different.  
Furthermore, we are aware that doping, i.e., varying $\mu$, strongly
affects the value of $T_K$ which may continuously vanish at the Dirac
point as we discussed in section~\ref{section:Kondotemperature}.  Nevertheless, 
we expect that the Friedel oscillations will
qualitatively not really depend on $T_K$. 
This is a consequence of the
universality of the strong Kondo coupling effective regime that is
realized below $T_K$ within the renormalization group picture: since
Friedel oscillations appear above the Kondo screening size $R_K$ their
shape is qualitatively universal (but still depends on the warping
length $R_W$ and Fermi pseudo period $\pi R_F$).  Also, according to the values given in 
table~\ref{Table:Bi2Se3}, the crossover value for the chemical potential varies very smoothly from 
$\mu^\star=160$~meV to $\mu^\star=400$~meV when $T_K$ changes from $10$~K to $1000$~K. 
This suggests that the results which are illustrated by figure~\ref{fig:intensityBi2Se3} for 
$T_K=100$~K can be extended to any other values of $T_K$. Of course, the characteristic unit length which is used for the plots, $R_K$, would have to be rescaled accordingly. Experimentally, one of the main difficulties for 
observing Friedel oscillations with or without six-fold symmetry is the requirement of cooling the temperature 
sufficiently lower than $T_K$, but the orders of magnitudes that are considered here correspond to values that are 
realistic both physically ($T_K$ is imposed by the chemistry) and technologically ($T$ is limited by cryogenic technics). 

We made a similar analysis for $\mathrm{Bi_2Te_3}$ compounds. In this case, 
Ref.~\onlinecite{chen2009} indicates experimental values of $\mu$
between $350$~meV and $120$~meV. We thus plotted these two extreme
cases, together with the intermediate value $\mu=250$~meV. Here, we chose a Kondo temperature 
$T_K=100$~K, which corresponds to a crossover value $\mu^\star=130$~meV as indicated in 
table~\ref{Table:Bi2Te3}. In this case, the six-fold symmetry resulting from the warping is thus observable for 
$\mu=350$~meV and $250$~meV, and the full rotation symmetry is recovered for $\mu=120$~meV. 
The six-fold symmetry may remain for that value of chemical potential if the Kondo temperature is lowered. 
Indeed, table~\ref{Table:Bi2Te3} indicates $\mu^\star=82$~meV for $\mathrm{Bi_2Te_3}$ compounds 
with $T_K=10$~K. 

Here, we have restricted our analysis to the observation of Friedel oscillations within the fluctuation of the local 
density of states, $\delta\rho({\bf r})$. This physical quantity can be measured experimentally using Scanning Tunneling 
Microscopy (STM). Local Density of States\cite{mitchell2012} (LDOS)
measurements have 
already been performed by STM on Kondo
impurities
at the surface of
metals.\cite{madhavan1998,madhavan2001,knorr2002,wahl2005,ysfu2007}  The measurement of the Friedel oscillations
would require the integration of the measured local density of states over a
range of energy\cite{affleck2008}. 
Beside the issue of cooling the temperature sufficiently lower than $T_K$, other technical limitations 
have to be considered in order to observe the predicted Friedel oscillations using STM: \\
First, a voltage bias is applied locally between the tip of the STM and the surface of the sample. The resulting STM 
current which is measured may invoke out of equilibrium effects that have not been analyzed here. We expect our 
predictions to be valid for STM experiments with bias voltages invoking energies that are lower to both $T_K$ and 
$\mu$. Higher values of bias voltage may have non universal effects on
the Kondo screening leading to a distortion of the Friedel oscillations. \\
A second limitation is the STM resolution in both lateral and depth directions. More precisely, we may expect 
an experimental STM measurement of the Friedel oscillations to be realized by moving the STM tip at the surface of the system along two orthogonal directions. The most natural resulting STM signal will thus be discretized on a grid having 
a square lattice symmetry and an elementary step of length
$R_{STM}\simeq 0.2$nm. Since the period of the Friedel oscillations is 
$\pi R_F$, the measured STM signal may exhibit a Moir\'e pattern\cite{kuwabara1990} resulting from the interference between the two periods, 
$R_{STM}$ and $\pi R_F$. Considering the values of $R_F$ which are given in tables~\ref{Table:Bi2Se3} 
and~\ref{Table:Bi2Te3}, and assuming $R_{STM}$ is of the order of one or few \AA, Moir\'e patterns might 
occur for values of chemical potential relatively higher than
$\mu^\star$. In such cases, the measured STM images  would only have
the two-fold symmetry common to both the square and the hexagonal
symmetry groups.  

Comparing qualitatively the plots of figures~\ref{fig:intensityBi2Se3} and~\ref{fig:intensityBi2Te3}, we find that warping 
effects and their related six-fold symmetry are more observable at the surface of Bi$_2$Te$_3$ rather than Bi$_2$Se$_3$. 
This is due of course to a larger value of the warping constant $\lambda$, but this also results from a smaller value of 
the Fermi velocity $v_F$, which gives a smaller value of crossover potential $\mu^\star$.


\subsection{Conclusion}
\label{sec:ccl}

We have shown that a magnetic impurity on the surface of a strong topological
insulator will be fully screened by the surface modes unless the Fermi energy
is exactly at the Dirac point. The result depends only on the time
reversal invariance of the effective Hamiltonian of the surface modes
and is valid in particular in the presence of warping of the Fermi
surface.  We have shown
that Friedel oscillations are formed around the impurity and we have
calculated the shape of these oscillations both  without
warping and with a weak warping that can be treated perturbatively. 
With warping, the symmetry of the Friedel oscillation pattern is
broken from full rotational symmetry to a six-fold symmetry. 
In both cases, the pseudo period of the oscillations, $\pi R_F=\pi v_F/\mu$, is half the Fermi wave length of the 
surface modes, the short distance cut-off $R_K\propto v_F/T_K$ is determined by the Kondo temperature $T_K$, 
and the long distance cut-off $R_T=v_F/T$ results from thermal fluctuations. 
With warping, the amplitude of the fully rotationally symmetric part decreases 
as $1/r^3$, whilst the 6-fold symmetry term has an envelope decreasing more slowly, 
a $1/r^2$. As a consequence, a new length scale $R_W$ emerges, above which Friedel oscillations with  
six-fold symmetry may be observed. The crossover condition $R_K\approx R_W$ defines a chemical potential 
$\mu^\star$ associated to a doping $n_{S}^\star$ above which the Friedel oscillations are characterized by the 
six-fold symmetry even at shortest distances. 
Considering realistic values for the model parameters, we analyzed the observability and the symmetry of 
Friedel oscillations in the vicinity of magnetic impurities deposited at the surface of two compounds, 
$\mathrm{Bi_2Se_3}$ and $\mathrm{Bi_2Te_3}$. We identified large range of parameters where the crossover between 
the 6-fold and the fully rotational symmetries may be observed. We propose to use STM as an experimental probe 
for the variation of the local density of states. 

Various questions remain to be addressed. First, it would be
interesting
to investigate the unstable fixed point that separates the regime
with Kondo screening from the regime of decoupled impurities in the
case of a system at half-filling. 
Second, an exact calculation of the Friedel oscillations could be
performed using form factor expansion
methods\cite{mussardo_offcritical_review,saleur_cambridge} at zero
temperature.  

The single impurity model that we considered here could also be generalized to 
two or several impurities. In this case, the local Kondo screening will compete with 
the Ruderman-Kittel-Kasuya-Yosida\cite{ruderman1954,kasuya1956,yosida1957} (RKKY) inter-impurity screening
as discussed by Doniach\cite{doniach1977} in a general context. 
Here, we may expect the 6-fold symmetry of the Friedel oscillations 
to have signatures on the symmetry of the RKKY interaction.

\begin{acknowledgments}
  We thank J. Cayssol, D. Carpentier, P. Coleman, N. Andrei,
  N. Perkins and  P. Simon for discussions. 
  The present work was supported by Agence Nationale de la
  Recherche under grant ANR 2010-BLANC-041902 (ISOTOP). 
\end{acknowledgments}

\appendix 
\section{Details of the mean-field approximation}\label{sec:derivation-t-matrix}
We start from the mean-field self-consistent relations~(\ref{eq:mf-general}): 
\begin{eqnarray}
\label{eq:mf-Delta-appendix}
|\Delta|^2 &=&  J_K  \int_{-\infty}^\infty \frac{dE}{\pi} 
  \frac{(E +\mu_f) n_F(E)\Sigma''(E)}{[E + \mu_f
    -\Sigma'(E)]^2 + [\Sigma''(E)]^2}~, \\
\label{eq:mf-muf-appendix}
  \frac 1 2 &=&-\int_{-\infty}^\infty \frac{dE}{\pi}
  \frac{ n_F(E)\Sigma''(E)}{[E+\mu_f-\Sigma'(E)]^2 + [\Sigma''(E)]^2}~, 
\end{eqnarray}
with the expression~(\ref{eq:Sigmasecond}) of the imaginary part of the self energy: 
\begin{eqnarray}
\Sigma''(E)&=&-\pi|\Delta|^2\rho_0(\mu+E)~. 
\end{eqnarray}
Introducing the Hilbert transform of the density of states: 
\begin{eqnarray}
\Phi(E)\equiv \int _{-\infty}^\infty d\epsilon\rho_0(\epsilon)
\mathrm{P.V.}\left( \frac{1}{E-\epsilon}\right)~,
\end{eqnarray}
the expression~(\ref{eq:Sigmaprime}) of the real part of the self-energy can be written as: 
\begin{eqnarray}
\Sigma'(E)&=&\pi |\Delta|^2 \Phi(\mu+E)~. 
\end{eqnarray}

\subsection{Derivation  of the Kondo temperature \label{Appendix:subsectionTK}}

When the temperature $T$ goes to $T_K$, the hybridization parameter $\Delta$
vanishes. In such limit, the Eq.~(\ref{eq:mf-muf-appendix}) reduces to 
 $n_F(-\mu_F)=1/2$, yielding $\mu_f(T_K)=0$. The 
Eq.~(\ref{eq:mf-Delta-appendix}) can then  be cast into the
Nagaoka-Suhl form\cite{nagaoka_resonance, suhl_resonance}:  
\begin{eqnarray} \label{eq:general-mf-TK} 
-\frac{1}{J_K}=\mathrm{P. V.} \int_{-\infty}^{+\infty} \frac{dE}{E}
n_F(E) \rho_0(\mu+E)  
\end{eqnarray}

In the general $\mu>0$ case, we can rewrite Eq.~(\ref{eq:general-mf-TK}):
\begin{eqnarray}
  - \frac 1 {J_K}&=&-\rho_0(\mu) \int_{0}^{D-\mu} \frac{dE}{2E} \tanh
  \left(\frac{E}{2T_K}\right) - \rho_0(\mu) \int_{D-\mu}^{D+\mu} \frac{dE}{E} n_F(-E) 
\nonumber \\ 
 && + \int_{-D -\mu}^{D-\mu} \frac{dE}{E} [\rho_0(\mu+E)-\rho_0(\mu)]
 n_F(E)~,  
\end{eqnarray}
where $D$ is a bandwidth cutoff such that $\rho(E)=0$ when $|E|>D$. 
taking the limit of low Kondo temperature $T_K\ll D,\mu$, we obtain the
following approximate equation: 
\begin{eqnarray}
\frac 1 {J_K}=\rho_0(\mu) \ln \left(\frac{2 e^{\gamma_E} (D+\mu)}{\pi
    T}\right) - \int_{-D-\mu}^0 \frac{dE}{E} [\rho_0(\mu+E)-\rho_0(\mu)]~,
\end{eqnarray}
improving the prefactor in the expression of the Kondo temperature by
taking into account the variation of the density of states with the energy. 

\subsection{Derivation of the resonance width}

We have seen that both $\Gamma$ and $\mu_f$ vanish at and above the Kondo temperature. 
Hereafter, we will consider the limit of small Kondo coupling, and we will thus assume that $\Gamma$ and $\mu_f$ 
remain small compared to the non-interacting electron characteristic energy scales, even below the Kondo temperature 
where these quantities are not vanishing any more. 
If we take first the equation (\ref{eq:mf-muf-appendix}), and make the
approximations $\Sigma'(E)=0$, $\Sigma"(E)=\Sigma"(0)=-\Gamma$, we obtain:
\begin{eqnarray}
  \frac 1 2 =\int_{-\infty}^\infty \frac{dE}{\pi}
  \frac{\Gamma}{(E+\mu_f)^2+\Gamma^2} n_F(E) \simeq n_F(-\mu_f)~,   
\end{eqnarray}
so we have to take also $\mu_f=0$. Inserting our approximations in the
denominator of (\ref{eq:mf-Delta-appendix}), we obtain a second
equation:
\begin{eqnarray}\label{eq:mf-gamma-appendix} 
  -\frac 1 {J_K}=\int_{-\infty}^\infty dE \frac{E n_F(E)
    \rho_0(\mu+E)}{E^2+\Gamma^2}~. 
\end{eqnarray}
We introduce the bandwidth cutoff $D$ such that
$\rho_0(E)=0$ if $|E|>D$. We can then rewrite
Eq.~(\ref{eq:mf-gamma-appendix}) in the form: 
\begin{eqnarray}
  -\frac 1 {J_K}=\rho_0(\mu) \int_{-D-\mu}^{D-\mu} dE \frac{E
      n_F(E)}{E^2+\Gamma^2} + \int_{-D-\mu}^{D-\mu} dE \frac{E n_F(E)
    [\rho_0(\mu+E)-\rho_0(\mu)]}{E^2+\Gamma^2}~.
\end{eqnarray}
In the rightmost integral of the right hand side,
$\rho(\mu+E)-\rho(\mu)$ vanishes for $E=0$, so we can neglect $\Gamma$
in the denominator. This gives the final equation: 
\begin{eqnarray}\label{eq:gamma-mf-final-appendix} 
  -\frac 1 {J_K\rho_0(\mu)}= \int_{-D-\mu}^{D-\mu} \frac{E
    n_F(E)}{E^2+\Gamma^2} dE + \int_{-D-\mu}^{D-\mu} 
  \frac{\rho(E+\mu)-\rho(\mu)}{\rho_0(\mu)}  n_F(E)  \frac{dE}{E}~.    
\end{eqnarray}
With that approximation, and 
taking the zero temperature limit, we obtain: 
\begin{eqnarray}
  \Gamma(T=0) =(D+\mu) \exp\left[-\frac 1 {J_K \rho_0(\mu)}
    -\int_{-D-\mu}^0 \frac {dE}{E}
    \frac{\rho_0(\mu+E)-\rho_0(\mu)}{\rho_0(\mu)}   \right]~.
\end{eqnarray}
We note that $\Gamma(T=0)/T_K=2e^{\gamma_E}/\pi\simeq 2.26$ for such
level of approximation. 
For finite temperature, we can replace $n_F(E)$ with $\theta(-E)$ in
the integral over the density of states in the right-hand side of
Eq.~(\ref{eq:gamma-mf-final-appendix}). Using that approximation, we
can write:
\begin{eqnarray}
  \label{eq:mf-gamma-T-appendix}
  \ln\left(\frac{\Gamma(T=0)}{\Gamma(T)}\right) =
  \int_{-\infty}^\infty \frac{du}{4\cosh^2 u} \ln \left(1+\frac{4 T^2
      u^2}{\Gamma^2}\right)~.  
\end{eqnarray}

These equations imply that
$\Gamma(T)=\Gamma(T=0)\varphi(\Gamma(T=0)/T)$, so given the relation
between $\Gamma(T=0)$ and $T_K$, $\Gamma(T)=\Gamma(T=0)\bar
\varphi(T_K/T)$.

\section{Asymptotic approximation for the Green's function in the
  presence of warping}\label{app:asymp}   

The exact Green's function of the surface electrons is: 
\begin{equation}
  \label{eq:exact-green}
  G_0(\mathbf{r},i\nu_n)=\int\frac{d^2\mathbf{k}}{(2\pi)^2}
  \frac{i\nu_n+\mu+
    v_F\mathbf{\hat{z}}\cdot(\boldsymbol{\sigma}\times\mathbf{k})+\frac{\lambda}{2}(k_+^3+k_-^2)\sigma^z}{(i\nu_n+\mu)^2-v_F^2 k^2-\left[\frac{\lambda}{2}(k_+^3+k_-^2)\right]^2} e^{i \mathbf{k}\cdot\mathbf{r}}~. 
\end{equation}

Using polar coordinates, we can express the Green's
function~(\ref{eq:exact-green}) as a series:
\begin{eqnarray}
  \label{eq:exact-gf-series}
  G_0(r,\theta,i\nu_n)&=&\int \frac{k dk}{(2\pi)^2} \int d\phi  \frac{i z + i v_F k(\sigma^+
    e^{-i\phi}-\sigma_-e^{i\phi}) + \lambda k^3 \cos (3\phi )
    \sigma^z}{\sqrt{(z^2 + v_F^2 k^2)(z^2   + v_F^2 k^2 +
      \lambda^2 k^6)}}e^{i k
    r \cos (\phi-\theta )}\times \nonumber \\ &&  \sum_{m=-\infty}^\infty \left(\frac{\lambda^2
      k^6}{(\sqrt{(z^2 + v_F^2 k^2)} +\sqrt{(z^2   + v_F^2 k^2 +
      \lambda^2 k^6)})^2}\right)^{|m|} (-e^{i 6\phi})^m~,  
\end{eqnarray}
where $z=\nu_n-i\mu$. 
To obtain an asymptotic expansion of $G(r,\theta,i\nu_n)$ to lowest
order in $\lambda$ it is enough to consider the term $m=0$. We then
have to consider the integral: 
\begin{eqnarray}
  \label{eq:gf-lowest-order}
  G_0(r,\theta,i\nu_n)=\int \frac{k dk}{(2\pi)^2} \int d\phi \frac{i z + i v_F k(\sigma^+
    e^{-i\phi}-\sigma_-e^{i\phi}) + \lambda k^3 \cos (3\phi )
    \sigma^z}{\sqrt{(z^2 + v_F^2 k^2)(z^2   + v_F^2 k^2 +
      \lambda^2 k^6)}}e^{i k
    r \cos (\phi-\theta)}+\cdots~, 
\end{eqnarray}
where $\cdots$ denotes corrections of higher order in $\lambda$. 
After integration with respect to $\phi$ this expression gives: 
\begin{eqnarray}
 G_0(r,\theta,i\nu_n)&=&\int_{0}^{\infty}dk 
\frac{izkJ_0(kr)
-v_Fk^2(\sigma^+e^{-i\theta}-\sigma_-e^{i\theta})J_1(kr)
-i \lambda k^4 \cos (3\theta)\sigma^zJ_3(kr)
}{2\pi\sqrt{(z^2 + v_F^2 k^2)(z^2+v_F^2 k^2 +\lambda^2 k^6)}}\\
&~&\nonumber\\
&\equiv&
\frac{iz}{2\pi}I
-\frac{v_F}{2\pi}(\sigma^+e^{-i\theta}-\sigma_-e^{i\theta})I'
-\frac{i \lambda}{2\pi}\cos (3\theta)\sigma^zI''~. 
\label{eq:GreenfunctionIIprimeIseconde}
\end{eqnarray}
In the limit $\lambda \to 0$, we have from Eq.~(11.4.44) in
Ref.~\onlinecite{abramowitz_math_functions}:  
\begin{eqnarray}
  \label{eq:lintegralsI}
I&\to&\int_0^\infty dk\frac{k J_0(kr)}{z^2 + v_F^2 k^2}=\frac 1
  {v_F^2} K_0\left(\frac{|\nu_n|-i \mu~\sign(\nu_n)}{v_F} r \right)~,  \\ 
I'&\to&\int_0^\infty dk\frac{k^2 J_1(kr)}{z^2 + v_F^2 k^2}=\frac {|\nu_n|-i \mu~\sign(\nu_n)}
  {v_F^3} K_1\left(\frac{|\nu_n|-i \mu~\sign(\nu_n)}{v_F} r \right)~,  
 \label{eq:lintegralsIprime}
\end{eqnarray}
with corrections of order $\lambda^2$. 
We are thus left with the evaluation of the integral: 
\begin{eqnarray}
\label{eq:remain-int}
I''=\int_0^{+\infty} \frac{k^4 J_3(kr) dk}{\sqrt{(z^2+(v_F
      k)^2)(z^2 +(v_F
      k)^2 +\lambda^2 k^6)}}~.
\end{eqnarray}
In the expression~(\ref{eq:remain-int}) we cannot take the limit
$\lambda \to 0$ directly before integrating as we would obtain a divergent integral. Instead, we
will first make integrations by parts using Eq.~(9.1.30) in
Ref.~\onlinecite{abramowitz_math_functions} to obtain an asymptotic
expansion. 
The first integration by parts gives:
\begin{eqnarray}
  I''&=&\frac 1 r \int_0^{+\infty} dk k^4 J_4(kr) \left[\frac{v_F^2 k}
    {(z^2+v_F^2 k^2)^{3/2}(z^2+v_F^2 k^2+ \lambda^2 k^6)^{1/2}} +\frac{v_F^2 k
    + 3 \lambda^2 k^5} 
    {(z^2+v_F^2 k^2)^{1/2}(z^2+v_F^2 k^2+ \lambda^2 k^6)^{3/2}} \right]~. \nonumber\\
&&
\label{eq:annexe-Iseconde}
\\
&=&
\frac{v_F^2}{r} \int_0^{+\infty} dk k^5 J_4(kr) \left[\frac{1}
    {(z^2+v_F^2 k^2)^{3/2}(z^2+v_F^2 k^2+ \lambda^2 k^6)^{1/2}} +\frac{1} 
    {(z^2+v_F^2 k^2)^{1/2}(z^2+v_F^2 k^2+ \lambda^2 k^6)^{3/2}}
  \right]+\cdots~, \nonumber\\
&&
\end{eqnarray}
where $\cdots$ denotes the third term in~(\ref{eq:annexe-Iseconde}), which will be shown to be  negligible in the limit 
$\lambda\to 0$ in App.~\ref{app:remainders}.   
Using another integration by parts, we thus have:
\begin{eqnarray}
I''&\simeq&\frac{v_F^2}{r^2} \int_0^{+\infty} k^5 J_5(kr) \left[\frac{3 v_F^2
    k}{(z^2 +v_F^2 k^2)^{5/2}(z^2 +v_F^2 k^2 + \lambda^2 k^6)^{1/2}}+\frac{ v_F^2
    k+3\lambda^2 k^5 }{(z^2 +v_F^2 k^2)^{3/2}(z^2 +v_F^2 k^2 + \lambda^2
    k^6)^{3/2}}    \right] \nonumber \\ 
&& + \frac{v_F^2}{r^2} \int_0^{+\infty} k^5 J_5(kr) \left[\frac{ v_F^2
    k}{(z^2 +v_F^2 k^2)^{3/2}(z^2 +v_F^2 k^2 + \lambda^2 k^6)^{3/2}}+\frac{ 3(v_F^2
    k+3\lambda^2 k^5) }{(z^2 +v_F^2 k^2)^{1/2}(z^2 +v_F^2 k^2 + \lambda^2
    k^6)^{5/2}}    \right] 
\label{eq:annexe-Iseconde-bp}  
\end{eqnarray}
If we take the limit $\lambda \to 0$ (justification for this can be found in App.~\ref{app:remainders}) in the above expression, we 
obtain a convergent integral:
\begin{equation}
  \label{eq:asym-warp}
I''\to   \frac{8 v_F^4}{r^2} \int_0^{+\infty} \frac{k^6 J_5(kr)}{(z^2 +
    v_F^2 k^2)^3} = \frac{(|\nu_n|-i \mu~\sign(\nu_n))^3}{v_F^5} K_3\left(
    \frac{|\nu_n|-i \mu~\sign(\nu_n)}{v_F} r\right)~.  
\end{equation}
We can check that the terms proportional to $\lambda^2$ give
contributions that are vanishing in the limit $\lambda \to 0$. 

Invoking the asymptotic expressions~(\ref{eq:lintegralsI}),~(\ref{eq:lintegralsIprime}), 
and~(\ref{eq:asym-warp}) together in Eq.~(\ref{eq:GreenfunctionIIprimeIseconde}), 
we obtain the following approximation
for the Green's function in the limit of small warping: 
\begin{eqnarray}
  \label{eq:green-asymp-warp}
  G(r,\theta,i\nu_n) &\simeq& \frac{i\nu_n +\mu}{2\pi
    v_F^2}\left[K_0\left(\frac{|\nu_n|-i \mu~\sign(\nu_n)}{v_F} r
    \right)+i  (\sigma^+ e^{-i \theta} -\sigma^- e^{i\theta})
   K_1\left(\frac{|\nu_n|-i \mu~\sign(\nu_n)}{v_F} r
    \right) \sign(\nu_n) \right. \nonumber \\ && +\left. \frac{\lambda}{v_F}\sigma^z
    \left(\frac{i\nu_n + \mu}{v_F}\right)^2 K_3\left(\frac{|\nu_n|-i \mu~\sign(\nu_n)}{v_F} r
    \right) \sign(\nu_n) \cos 3\theta \right]~. 
\end{eqnarray}

\section{Evaluation of the remainders} 
\label{app:remainders}

We consider the integral giving the remainder in Eq.(\ref{eq:annexe-Iseconde}):  
\begin{eqnarray}
  I_3=\frac{3\lambda^2}{r}\int_0^{+\infty} dk \frac{k^9 J_4(kr)}{(z^2+(vk)^2)^{1/2} (z^2+(vk)^2+\lambda^2 k^6)^{3/2}}. 
\end{eqnarray}
With the change of variables $k=\sqrt{v_F/\lambda} u$, the integral $I_3$ is rewritten:
\begin{eqnarray}
  I_3=\frac{3}{\lambda v_F r} \int_0^{+\infty} du \frac{ u^9 J_4(au)}{(b^2 + u^2)^{1/2} (b^2+u^2+u^6)^{3/2}},  
\end{eqnarray}
with $a=r(v_F/\lambda)^{1/2}$ and $b^2=\lambda z^2/v_F^3$. 
To obtain an upper bound for the integral, we use 3 successive integration by parts to rewrite: 
\begin{eqnarray}
\label{eq:by-parts-3} 
  I_3=\frac{3}{ a^3 \lambda v_F r}\int_0^{+\infty} du \frac{u^7 J_7(ua) P(u)}{(b^2+u^2)^{7/2} (b^2+u^2+u^6)^{9/2}},  
\end{eqnarray}
 where the polynomial $P(u)$ has the expression:   
\begin{eqnarray}
  \label{eq:polynomial}
  P(u)&=&480\,u^{22}+1200\,b^2\,u^{20}+1050\,b^4\,u^{18}-360\,u^{18}+315\,b^
 6\,u^{16}-3276\,b^2\,u^{16}-7200\,b^4\,u^{14}-6084\,b^6\,u^{12}-432
 \,b^2\,u^{12} \nonumber \\ && -1800\,b^8\,u^{10}-432\,b^4\,u^{10}+1152\,b^6\,u^8-96\,
 b^2\,u^8+1872\,b^8\,u^6-192\,b^4\,u^6+720\,b^{10}\,u^4+192\,b^8\,u^2
 +96\,b^{10}~. 
\end{eqnarray}

The numerator in Eq.~(\ref{eq:by-parts-3}) is $O(u^{57/2})$ while the denominator is $O(u^{34})$ for $u\to \infty$ making the integral in (\ref{eq:by-parts-3}) convergent. Moreover, an upper bound for the integral is given by: 
\begin{eqnarray}
  \int_0^{+\infty} du \frac{|P(u)|}{\sqrt{2} (b^2+u^2+u^6)^{9/2}}~,   
\end{eqnarray}
where we have used the inequalities\cite{abramowitz_math_functions} $|J_7(u)|<1/\sqrt{2}$ and $(b^2+u^2)^{-7/2}<u^{-7}$.We can them majorize the polynomial $P(u)$ by the sum of the absolute value of its monomials. For monomials of degree $n\ge 9$, we can also use the inequality $(b^2+u^2+u^6)^{9/2}> u^9 (1+u^4)$ to obtain an upper bound larger than $b$. For the monomials of degree $n<8$, we obtain an upper bound of the form:
\begin{eqnarray}
  b^m \int_0^{+\infty} \frac{u^n du}{(b^2+u^2+u^6)^{9/2}} <  b^m \int_0^{+\infty} \frac{u^n du}{(b^2+u^2)^{9/2}}=b^{n+m-8} \int_0^{+\infty} \frac{u^n du}{(1+u^2)^{9/2}}~.  
\end{eqnarray}
 From the expression~(\ref{eq:polynomial}) of $P(u)$, we see that these terms contribute expressions $O(b^2)$.  
In the case of $n=8$, we have to consider:
\begin{eqnarray}
   \int_0^{+\infty} \frac{u^n du}{(b^2+u^2+u^6)^{9/2}} \le \int_0^{1/b} du \frac{u^8}{(1+u^2)^{9/2}} + \int_1^{+\infty} \frac{du}{u (1+u^4)^{9/2}}~. 
\end{eqnarray}
Inspecting the polynomial $P(u)$, we see that the contribution of the $u^8$ terms will be $O(b^2 |\ln b|)$. 
Putting all contributions together, we see that:
\begin{eqnarray}
  I_3 &\le& \frac {3}{a^3 \lambda v_F r} (C+ O(b^2 |ln b|)) \\
      &\le&  \frac{3}{v_F^2 r^3} \sqrt{\frac{\lambda}{v_F}} (C+ O(\lambda |\ln \lambda|)~,  
\end{eqnarray}

So $I_3=O(\lambda^{1/2})$ for $\lambda \to 0$. This establishes that this term gives a subdominant contribution to the Matsubara Green's function.

We can apply the same method to the remainder integrals appearing in Eq. (\ref{eq:annexe-Iseconde-bp}). 

If we consider the integral:
\begin{eqnarray}
  I_4= \int_0^\infty \frac{\lambda^2 k^{10} J_5(kr) }{(z^2 +v_F^2 k^2)^{3/2}(z^2 +v_F^2 k^2 + \lambda^2
    k^6)^{3/2}}~,  
\end{eqnarray}
by the same change of variables $k=(v/\lambda)^{1/2}u$, we can transform it into: 
\begin{eqnarray}
  I_4=\frac 1 {v_F^{7/2} \lambda^{1/2}} \int_0^{+\infty} \frac{u^{10} J_5(ua)}{(b^2+u^2)^{3/2} (b^2+u^2+u^6)^{3/2}}~.  
\end{eqnarray}

Integrating by parts twice, we rewrite: 
\begin{eqnarray}
  I_4=\frac 1 {v_F^{7/2} a^2 \lambda^{1/2}} \int_0^{\infty} du \frac{u^7 J_7(ua) P_4(u)}{(b^2+u^2)^{7/2} (b^2+u^2+u^6)^{7/2}}~, 
\end{eqnarray}
where:
\begin{eqnarray}
  P_4(u)&=&80\,u^{17}+100\,b^2\,u^{15}+35\,b^4\,u^{13}+28\,u^{13}-84\,b^2\,u^{
 11}-204\,b^4\,u^9+8\,u^9-92\,b^6\,u^7-16\,b^2\,u^7-48\,b^4\,u^5\nonumber \\ && -16\,
 b^6\,u^3+8\,b^8\,u~. 
\end{eqnarray}
We can show that the integral remains finite in the limit of $\lambda \to 0$, and as a result, $I_4=O(\lambda^{1/2})$. 

Similarly, for the integral: 
\begin{eqnarray}
 I_5=\int_0^{+\infty} \frac{ \lambda^2 k^{10} J_5(kr) }{(z^2 +v_F^2 k^2)^{1/2}(z^2 +v_F^2 k^2 + \lambda^2
    k^6)^{5/2}}~, 
\end{eqnarray}
 we can rewrite:
 \begin{eqnarray}
   I_5=\frac 1 {v_F^{7/2} \lambda^{1/2}} \int_0^{+\infty} \frac{u^{10} J_5(ua)}{(b^2+u^2)^{1/2} (b^2+u^2+u^6)^{5/2}}~,  
 \end{eqnarray}
and by integrating by parts twice, we find: 
\begin{eqnarray}
  I_5=\frac 1 { a^2 v_F^{7/2} \lambda^{1/2}}  \int_0^{+\infty} \frac{u^{7} J_7(ua) P_5(u)}{(b^2+u^2)^{5/2} (b^2+u^2+u^6)^{9/2}}~,  
\end{eqnarray}
where: 
\begin{eqnarray}
P_5(u)&=& 168\,u^{17}+308\,b^2\,u^{15}+143\,b^4\,u^{13}+36\,u^{13}-108\,b^2\,
 u^{11}-308\,b^4\,u^9+8\,u^9-164\,b^6\,u^7-16\,b^2\,u^7-48\,b^4\,u^5 \nonumber \\ &&
 - 16\,b^6\,u^3+8\,b^8\,u~,  
\end{eqnarray}
Repeating the previous reasoning, we again establish that $I_5=O(\lambda^{1/2})$. 

It should be noted that the integrations by part can be iterated as long as the resulting integrals have a finite upper bound for small $\lambda$.  This implies that our $O(\sqrt{\lambda})$  estimate for  $I_3$, $I_4$ and $I_5$ is only a conservative one. If the integrations by part can be repeated indefinitely, the result of the process is that $I_{3,4,5}=O(\lambda^n)$ for any $n>0$, a hint that the integrals may actually vanish with an essential singularity in the limit of $\lambda \to 0$.

\end{document}